\documentclass[pre,reprint,aps,superscriptaddress,showpacs,floatfix]{revtex4-2}
\usepackage[dvipdfmx]{graphicx}
\usepackage{bm,color,amsmath,amssymb,multirow,CJK,grffile}
\usepackage{mathrsfs}
\usepackage[title]{appendix}
\usepackage{braket}
\newcommand{\Tensor}[1]{\overset{\text{\tiny$\leftrightarrow$}}{#1}}
\newcommand{\R}[1]{{\mathrm{#1}}}
\newcommand{\F}[2]{\frac{#1}{#2}}

\newcommand{\K}[1]{\left(#1 \right)}
\newcommand{\BK}[1]{\left[#1 \right]}
\newcommand{\OD}[2]{\frac{d #1}{d #2}}
\newcommand{\PD}[2]{\frac{\partial #1}{\partial #2}}

\newcommand{\CK}[1]{\left\{#1 \right\}}
\newcommand{\ABS}[1]{\left| #1 \right|}

\newcommand{\MYBLACK}[1]{\color{black}#1 \color{black}}
\begin{document}
\begin{CJK*}{UTF8}{}
	\title{
		Oscillating edge current in polar active fluid
		}

		\author{Hiroki Matsukiyo
		}
		\affiliation{
			Department of Physics, Kyushu University,
			744 Motooka, Nishi-ku, Fukuoka 819-0395, Japan
			}
			\author{Jun-ichi Fukuda
			}
			\affiliation{
				Department of Physics, Kyushu University,
				744 Motooka, Nishi-ku, Fukuoka 819-0395, Japan
				}
				\date{\today}
				\begin{abstract}
					Dense bacterial suspensions exhibit turbulent behaviour called ``bacterial turbulence''.
					The behavior of the bulk unconstrained bacterial turbulence is described well by the Toner-Tu-Swift-Hohenberg (TTSH) equation
					for the velocity field.
					However, it remains unclear how we should treat boundary conditions on bacterial turbulence in contact with some boundaries (e.g. solid walls).
					To be more specific, although the importance of the ``edge current'', the flow along the boundary,
					has been demonstrated in several experimental
					studies on confined bacterial suspensions,
					previous numerical studies based on the TTSH equation employ non-slip boundary conditions
					and do not seem to describe properly the behavior of bacteria near the boundaries.
					In this study, we impose a slip boundary condition on the TTSH equation to describe the bacterial motion at boundaries.
					We develop a method to implement the slip boundary condition.
					Using this method, we have successfully produced edge current and discovered that the direction of the edge current temporally oscillates.
					The oscillation can be attributable to the advection term in the TTSH equation.
					Our work demonstrates that boundary conditions could play an important role in the collective dynamics of active systems.
				\end{abstract}
				\maketitle
\end{CJK*}
\section{Introduction}
\label{introduction}
Turbulence-like behaviour is observed in a wide range of active matter systems. It is called ``active turbulence'' (for an inclusive review, see ref.\cite{alert2022active}).
As listed in ref.\cite{alert2022active},
active turbulence has been reported in various kinds of experimental systems, such as sperm suspensions \cite{creppy2015turbulence},
self-propelled Janus particles \cite{nishiguchi2015mesoscopic}, tissue cell monolayers \cite{lin2021energetics}, microtubule-kinesin
suspensions \cite{guillamat2017taming,martinez2021scaling}, etc.

Dense bacterial suspensions also exhibit turbulent behavior called ``bacterial turbulence'' \cite{alert2022active,Wensink,Dunkel}.
Bacterial turbulence has a characteristic size of vortex, larger than the size of each single bacterium, and then exhibits a peaked energy spectrum.
Such characteristic velocity/vorticity profiles and energy spectrum are well-described by
the Toner-Tu-Swift-Hohenberg (TTSH) equation
\cite{Wensink,Dunkel,alert2022active}
	in which the coarse-grained collective velocity field $\bm{v}$, which is given by the sum of the velocity of the solvent fluid and that of the swimmers with respect to the fluid \cite{HeidenreichPRE2016,ReinkenPRE2018}, is considered as the only field variable describing the state of the system, and other degrees of freedom
(polar order, orientational order, density) are not taken into account (for example, orientational order is assumed to be parallel to $\bm{v}$).
The TTSH equation is given by
\begin{equation}
	\label{TTSH_eq_fluid1}
	\begin{split}
		\K{\partial_t+\lambda_0\bm{v}\cdot\bm{\nabla}}\bm{v}=&-\bm{\nabla}p+\lambda_1\bm{\nabla}\K{\bm{v}^2}\\
		&-\K{\alpha+\beta\ABS{\bm{v}}^2}\bm{v}\\
		&+\Gamma_0\bm{\nabla}^2\bm{v}-\Gamma_2\K{\bm{\nabla}^2}^2\bm{v}
	\end{split}
\end{equation}
and the incompressibility condition
\begin{equation}
	\label{incompressible}
	\bm{\nabla}\cdot\bm{v}=0.
\end{equation}
In eq.(\ref{TTSH_eq_fluid1}), $\lambda_0$, $\lambda_1$, $\alpha$, $\beta$, $\Gamma_0$ and $\Gamma_2$ are constants determined phenomenologically
and $p$ is the Lagrange multiplier to ensure the incompressibility (eq.(\ref{incompressible})).
The TTSH model does not explicitly take account of the solvent in which bacteria swim \cite{alert2022active}, 
and therefore breaks the Galilean invariance, which is reflected in the presence of
the $\lambda_0$ $\K{\ne 1}$, $\lambda_1$, $\alpha$ and $\beta$-terms.
The term $ -\K{\alpha+\beta\ABS{\bm{v}}^2}\bm{v} $, called the
Toner-Tu term, gives a characteristic speed $v_0=\sqrt{\ABS{\alpha}/\beta}$ and the
term $ \Gamma_0\bm{\nabla}^2\bm{v}-\Gamma_2\K{\bm{\nabla}^2}^2\bm{v} $, called the Swift-Hohenberg term,
gives a characteristic length scale $\Lambda_0=2\pi\sqrt{2\Gamma_2/\ABS{\Gamma_0}}$.

In recent years, ``control'' of the bacterial turbulence has been attracting interest.
More specifically, by imposing geometrical confinements on bacterial suspensions 
(e.g. confining bacterial suspensions in microscopic devices, locating small obstacles in bacterial suspensions etc.),
it has been shown that the seemingly chaotic motions of bacterial turbulence can be rectified in many experimental systems:
Wioland et al. \cite{wioland2016ferromagnetic} confined bacterial suspensions into chambers connected by channels and demonstrated the transition between
ferromagnetic and antiferromagnetic vortex order (where the directions of the adjacent vortices are same and different, respectively) by
varying the width of channels.
Beppu et al. \cite{beppu2017geometry} performed experiments using dumbbell-shaped devices and showed ferro-antiferromagnetic vortex order transition
by varying the distance between the centers of the two circles.
Nishiguchi et al. \cite{nishiguchi2018engineering} realized antiferromagnetic vortex order by locating, in bacterial suspensions, periodic arrays of
microscopic vertical pillars whose lateral size is comparable with the length of a single bacterium.

To study such systems numerically, it is natural to apply the TTSH equation to the situations where some boundaries exist.
There already exist several studies along this direction:
Reinken et al. \cite{Reinken_2020,Reinken_2022} suggested a numerical method to implement many small obstacles located in bacterial turbulence and
obtained the results consistent with experiments.
Puggioni et al. \cite{puggioni2022giant,puggioni2023flocking} performed TTSH simulations in confined circular
domains whose radius was much larger than the typical vortex size and showed emergence of a giant vortex whose size was comparable with the circular domains.
Shiratani et al. \cite{shiratani2023route} also performed simulations
in which the radius of the circular domain was varied with time,
and discovered a hysteresis of the transition between the single-vortex stationary state and the vortex-pair oscillatory state.

As stated above, the previous TTSH simulations revealed several interesting phenomena.
However, the simulation methods used there seem to fail to capture the behavior at boundaries because 
they cannot realize ``edge current'' i.e., the bacterial flow along the boundaries.

Edge current emerging in various active systems
(bacterial suspensions\cite{Beppu_2021}, active nematics\cite{Yamauchi_2020,Yashunsky_etal_PRX_2022}, active spinner materials\cite{vanZuiden_2016,Soni_2019}, etc.)
has been attracting interest recently.
In this paper, we particularly focus on bacterial suspensions described by the TTSH model.
Several experimental studies on confined bacterial suspensions demonstrate that the edge current
gives an essential effect on their collective motion \cite{wioland2013confinement,wioland2016ferromagnetic,beppu2017geometry,Beppu_2021}.
However, in all the above TTSH simulations\cite{Reinken_2020,Reinken_2022,puggioni2022giant,puggioni2023flocking,shiratani2023route}, a damping term is introduced
to represent the boundary and then non-slip boundary condition is imposed.
As stated clearly in ref.\cite{shiratani2023route},
in the simulations using non-slip boundary condition, the magnitude of the velocity continuously decays to zero at the boundaries due to the damping terms and
the edge current does not emerge.
Therefore a non-slip boundary condition does not describe properly the bacterial motion at the boundaries.

In this study, we
investigate how the difference of boundary conditions affects bacterial collective dynamics.
We propose an extended TTSH model to implement the slip boundary condition (Sec.\ref{model}) and
furthermore, present a numerical method to calculate our extended model equations (Sec.\ref{numerical_method}).
As a result of simulations, we have successfully realized the edge current (Sec.\ref{results_discussion}). Furthermore, we discovered that
the direction of the edge current temporally oscillates (Sec.\ref{results_discussion}).
Note that we consider the two-dimensional TTSH equation in this paper.

\section{Model}
\label{model}
To describe the bacterial dynamics, we use the Toner-Tu-Swift-Hohenberg (TTSH) equation \cite{Wensink,Dunkel,alert2022active}
(eq.(\ref{TTSH_eq_fluid1}), (\ref{incompressible})),
already mentioned in Sec.\ref{introduction}.
Note that the $ \lambda_1 $-term in eq.(\ref{TTSH_eq_fluid1}), can be absorbed into the $ p $-term:
$ -\bm{\nabla}p+\lambda_1\bm{\nabla}\K{\bm{v}^2} = -\bm{\nabla}\K{p-\lambda_1\bm{v}^2}\equiv -\bm{\nabla}q$, where we
have introduced a new Lagrange multiplier $ q $.
Let us use the TTSH equation without $ \lambda_1 $-term in the following discussions.

Next, for the following discussions, let us rewrite the TTSH equation (eq.(\ref{TTSH_eq_fluid1})) in terms of functional derivative:
\begin{equation}
	\label{TTSH_eq_fluid}
	\partial_{t}\bm{v}=-\F{\delta\mathcal{F}}{\delta\bm{v}}
	-\MYBLACK{\lambda_{0}\bm{v}\cdot\bm{\nabla}\bm{v},}
\end{equation}
where we have introduced the functional
\begin{equation}
	\begin{split}
		\label{free_energy}
		\mathcal{F}\BK{\bm v}=&\int d\bm{r}\left\{-q\partial_{i}v_i + \F{\alpha}{2}v_i v_i + \F{\beta}{4}v_i v_i v_j v_j \right. \\
		& \left.\hspace{30pt} +\F{\Gamma_0}{2}\K{\partial_j v_i}\K{\partial_j v_i} \right. \\
		& \left.\hspace{30pt} +\F{\Gamma_2}{2}\K{\partial_j \partial_j v_i}\K{\partial_k \partial_k v_i}
		\right\}. 
	\end{split}
\end{equation}
The first term in eq.(\ref{TTSH_eq_fluid}), $ -\delta\mathcal{F}/\delta\bm{v} $,
drives the system towards the minimization of the functional $ \mathcal{F} $.
The second term, $ -\lambda_{0}\bm{v}\cdot\bm{\nabla}\bm{v} $, is the advective term and $ \lambda_0 $ is the strength of the advection.

To implement the slip boundary condition, we replace the functional $ \mathcal{F} $ with $\mathscr{F}$ given by
\begin{equation}
	\label{eq:free_energy_functional_slip}
	\begin{split}
		\mathscr{F}\BK{\bm v}=&\int_{V} d\bm{r}
		\left\{
			-q\partial_iv_i+\F{\alpha}{2}v_iv_i+\F{\beta}{4}v_iv_iv_jv_j \right.\\
			&\hspace{30pt}\left. +\F{\Gamma_0}{2}\K{\partial_jv_i}\K{\partial_jv_i}\right.\\
			&\hspace{30pt}\left. +\F{\Gamma_2}{2}\K{\partial_j\partial_jv_i}\K{\partial_k\partial_kv_i}\right\}\\
		&+\F{\xi}{2}\int_{S}dS\K{{\bm n}\times{\bm v}}_{z}^2,
	\end{split}
\end{equation}
where the body integral $\int_{V} d\bm{r}$ is taken over the fluid region, the surface integral $\int_{S}dS$ is taken over the surface ($S$) of $V$
and $\bm{n}$ is the outward unit normal vector at the surface $S$.
The surface term $\F{\xi}{2}\int_{S}dS\K{{\bm n}\times{\bm v}}_{z}^2$ can be regarded as the energetic penalty for the slip velocity $\K{{\bm n}\times{\bm v}}_{z}$.
$\xi \K{>0}$ is the drag coefficient.
When $\xi =0$, no penalty arises for any finite slip velocity and the functional $ \mathscr{F}\BK{\bm v} $ gives stress-free slip boundary condition.
On the other hand, when $\xi\rightarrow\infty$, infinite penalty arises for arbitrary finite slip velocity and the functional $ \mathscr{F}\BK{\bm v} $ gives non-slip boundary condition
(see Appendix \ref{non_slip_limit}).
The boundary condition for a finite value of $\xi$ will be discussed in Sec.\ref{limit_d_zero} after specifying the shape of the boundary.
\section{Numerical Method}
\label{numerical_method}
\subsection{Smoothed Profile Method}
\label{spm}
To solve the extended model (eq.(\ref{TTSH_eq_fluid}) with the functional $\mathscr{F}\BK{\bm{v}}$, given by eq.(\ref{eq:free_energy_functional_slip})) numerically,
we use the smoothed profile method (SPM) \cite{Nakayama_Yamamoto,Kanke_Sasaki_2013},
where fluid-solid boundaries are represented by the smoothed profile $\phi$:
\begin{equation}
	\begin{split}
		\label{phi}
		\phi
		=\begin{cases}
			0 & \text{in solid regions}\\
			\text{varies between 0 and 1 smoothly} & \text{at boundaries}\\
			1 & \text{in fluid regions}.
		\end{cases}
	\end{split}
\end{equation}
Let us consider confining the bacterial suspensions in a closed circular domain.
To represent the circular domain whose radius is $ R $ and center is at the origin,
the three regions are defined as follows:
\begin{equation}
	\begin{split}
		\label{phi2}
		r<R-\delta &: \text{fluid}\\
		R-\delta\le r\le R+\delta &: \text{boundary}\\
		r>R+\delta &: \text{solid}
	\end{split}
\end{equation}
where $ 2\delta $ is the thickness of the smoothed boundary and $ r $ is the distance from the origin.
There are several candidates satisfying the above condition (eq.(\ref{phi}) and (\ref{phi2})).
Let us adopt the seemingly simplest one:
\begin{equation}
	\label{phi_tanh}
	\phi\K{r}=\F{1}{2}\tanh\F{R-r}{d}+\F{1}{2},
\end{equation}
where the thickness of the fluid-solid boundary is order $d$.
Note that the smoothed profile eq.(\ref{phi_tanh}) is not exactly equal to 0/1 in the solid/fluid regions and
we introduce a cut-off to divide the three regions (see Sec.\ref{practical_techniques} for details).

Using this smoothed profile $\phi$, we modify the integrals in the functional $ \mathscr{F}\BK{\bm v} $ (eq.(\ref{eq:free_energy_functional_slip})) as follows:
\begin{align}
	\label{eq:fluid_integral}
	\int_{V} d\bm{r}\quad &\rightarrow\quad\int d\bm{r}\phi,\\
	\label{eq:surface_integral}
	\int_{S}dS\quad &\rightarrow\quad\int d\bm{r}\ABS{\bm{\nabla}\phi},
\end{align}
where the integral $\int d\bm{r}$ is taken over the whole (fluid, solid and their boundary) regions.

Eq.(\ref{eq:free_energy_functional_slip}) is now rewritten as
\begin{equation}
	\label{eq:free_energy_phi}
	\begin{split}
		\mathscr{F}_{\phi}\BK{\bm v}=&\int d\bm{r}
		\phi\left\{
			-q\partial_iv_i+\F{\alpha}{2}v_iv_i+\F{\beta}{4}v_iv_iv_jv_j \right.\\
			&\hspace{35pt}\left. +\F{\Gamma_0}{2}\K{\partial_jv_i}\K{\partial_jv_i} \right.\\
			&\hspace{35pt}\left. +\F{\Gamma_2}{2}\K{\partial_j\partial_jv_i}\K{\partial_k\partial_kv_i}\right\}\\
		&+\F{\xi}{2}\int d\bm{r}\ABS{\bm{\nabla}\phi}
		\K{{\bm n}\times{\bm v}}_{z}^2.
	\end{split}
\end{equation}
Using the above functional $\mathscr{F}_{\phi}\BK{\bm{v}}$, the basic equation of our simulation is given by
\begin{equation}
	\label{TTSH_eq_fluid_phi}
	\partial_{t}\bm{v}=-\F{\delta\mathscr{F}_{\phi}\BK{\bm{v}}}{\delta\bm{v}}
	-\MYBLACK{\lambda_{0}\phi\bm{v}\cdot\bm{\nabla}\bm{v}.}
\end{equation}
Note that $ \phi $ is also put in the $ \lambda_0 $-term.
The above replacements using the smoothed profile $ \phi $ (i.e. eq.(\ref{eq:fluid_integral}), (\ref{eq:surface_integral}) and the $\lambda_0$-term in eq.(\ref{TTSH_eq_fluid_phi}))
will be justified in Sec.\ref{limit_d_zero} and Appendix.\ref{appendix}.
\subsection{The limit boundary thickness $\rightarrow 0$}
\label{limit_d_zero}
Let us identify the forms of our modified TTSH equation (eq.(\ref{TTSH_eq_fluid_phi}))
in each (solid, fluid and boundary) region
in the limit of $d\rightarrow0$.
We show here only the outline of the calculation and the results. For the details, see Appendix.\ref{appendix}.

In the limit $d\rightarrow0$, $\phi\K{\bm{r}}\rightarrow\Theta\K{R-r}$ and $\bm{\nabla}\phi\rightarrow-\delta\K{r-R}\bm{n}$, where
$\Theta\K{\cdot}$ is the step function and $\delta\K{\cdot}$ is the delta function.
In the fluid region, eq.(\ref{TTSH_eq_fluid_phi}) reduces to the TTSH equation with no boundary (eq.(\ref{TTSH_eq_fluid1})).
In the solid region, eq.(\ref{TTSH_eq_fluid_phi}) gives
\begin{equation}
	\label{equation_in_solid}
	\partial_t \bm{v}=0.
\end{equation}
Hence, by giving the initial condition where $\bm{v}=0$ in the solid region, $\bm{v}=0$ is satisfied in the subsequent time steps.

At the fluid-solid boundary, we obtain
\begin{equation}
	\label{equation_at_boundary}
	-\xi\K{\bm{n}\times\bm{v}}_z=\CK{\bm{n}\times\K{\Tensor{\sigma}\cdot\bm{n}-\Gamma_2\F{1}{R}\bm{\nabla}^2\bm{v}}}_z,
\end{equation}
where $\Tensor{\sigma}$ is the stress tensor of the TTSH equation defined by
\begin{equation}
	\partial_t v_i = \partial_j \sigma_{ij} -\K{\alpha+\beta v_jv_j}v_i,
\end{equation}
\begin{equation}
	\label{sigma_TTSH}
	\sigma_{ij}\equiv
	-q\delta_{ij}-\lambda_0v_iv_j
	+\Gamma_0\partial_jv_i-\Gamma_2\partial_j\partial_k\partial_kv_i.
\end{equation}
Eq.(\ref{equation_at_boundary}) is the Navier slip boundary condition
($-\xi\K{\bm{n}\times\bm{v}}_z=\BK{\bm{n}\times\K{\Tensor{\sigma}\cdot\bm{n}}}_z$)
with a correction term ($-\K{\Gamma_2/R}\K{\bm{n}\times\bm{\nabla}^2\bm{v}}_z$).
For more details on the Navier slip boundary condition,
see e.g. ref.\cite{neustupa2007navier}.
The Navier slip boundary condition
states that the tangential component of the velocity at the boundary is proportional to the tangential component of the stress and
has been used not only for Navier-Stokes fluids but also for several active systems
\cite{Yashunsky_etal_PRX_2022,Ramaswamy_SciRep_2016}.
\subsection{Remarks on the simulation techniques}
\label{practical_techniques}
Let us comment on the practical techniques to calculate eq.(\ref{TTSH_eq_fluid_phi}).
In our numerical simulations, eq.(\ref{TTSH_eq_fluid_phi}) is rewritten in terms of the stream function $\psi$, defined by $v_{i}=\epsilon_{ij}\partial_{j}\psi$, where
$ \epsilon_{ij} $ is the two-dimensional Levi-Civita symbol,
defined by $ \epsilon_{xy}=-\epsilon_{yx}=1 $, $ \epsilon_{xx}=\epsilon_{yy}=0 $
(for details, see Appendix.\ref{psi_representation}).
By introducing $\psi$, the incompressibility, eq.(\ref{incompressible}), is automatically satisfied and there is no need to
solve the Poisson equation for the Lagrange multiplier $q$.
From the stream function $\psi$, we can calculate the velocity $v_{i}\K{=\epsilon_{ij}\partial_{j}\psi}$, the vorticity $\omega\K{=-\bm{\nabla}^2 \psi}$ and their spatial derivatives.

We use a pseudo-spectral method and the fourth order Runge-Kutta formula for the discretization in space and time, respectively \cite{Canuto_1988}.
The cut-off wave number $K$ of the Fourier expansion is chosen to satisfy the 3/2-rule for $K$ and
the number of lattice points in each direction $J$: $J\ge 3K+1$ (see TABLE.\ref{table_fixed_parameters}).

The initial velocity field is random in the region $r<0.7R$ and 0 in the region $r>0.7R$.

As already mentioned in Sec.\ref{spm}, we introduce a cut-off to divide the three (solid, fluid and boundary) regions as follows:
\begin{equation}
	\begin{split}
		\label{cutoff_phi}
		r < R - 1.5d &: \text{fluid},\\
		R - 1.5d < r < R + 1.5d &: \text{boundary},\\
		R + 1.5d < r &: \text{solid},\\
	\end{split}
\end{equation}
where $ r $, $ R $ and $ d $ are the same as in eq.(\ref{phi_tanh}).

To satisfy the impermeability $ \bm{n}\cdot\bm{v}=0 $ at boundaries and $ \bm{v}=0 $ in the solid regions,
we set the normal component of the velocity at the boundary and the velocity in the solid region to zero at each time step.

In Sec.\ref{results_discussion},
all physical quantities are non-dimensionalized using the characteristic speed of bacteria $v_0$
and the characteristic length of spacial pattern $\Lambda_0$.
The non-dimensionalized values of the fixed parameters through all the simulations are presented
in TABLE.\ref{table_fixed_parameters}.
\begin{table}[btp]
	\caption{Fixed parameters through all the simulations. $v_0=\sqrt{\ABS{\alpha}/\beta}$ and $\Lambda_0=2\pi\sqrt{2\Gamma_2/\ABS{\Gamma_0}}$.}
	\label{table_fixed_parameters}
	\centering
	\begin{tabular}{p{60pt}c}
		\hline
		$J$ & $256$\\
		$K$ & $85$\\
		$\alpha/\K{v_0/\Lambda_0}$ & $-0.27$\\
		$\beta/\K{1/\K{v_0\Lambda_0}}$ & $0.27$\\
		$\Gamma_0/\K{v_0\Lambda_0}$ & $-0.078$\\
		$\Gamma_2/\K{v_0\Lambda_0^3}$ & $0.00099$\\
		$d/\Lambda_0$ & $0.31$\\
		\hline
	\end{tabular}
\end{table}
\section{Results and Discussion}
\label{results_discussion}
\subsection{Vorticity/velocity profiles and oscillation of edge current}
\begin{figure}[btp]
	\begin{center}
		\includegraphics[width=0.50\textwidth]{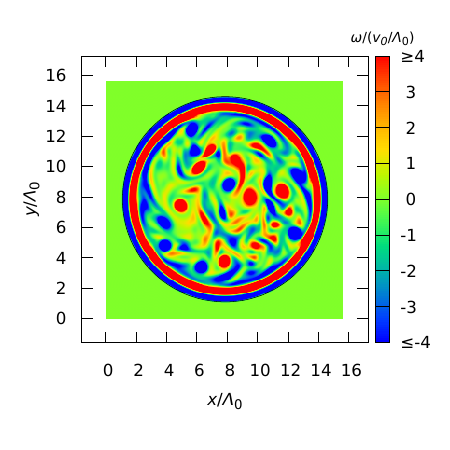}
	\end{center}
	\caption{
		A typical simulation snapshot of the vorticity field $ \omega/\K{v_0/\Lambda_0}=\K{\bm{\nabla}\times\bm{v}}_z/\K{v_0/\Lambda_0} $
		at the time $ t/\K{\Lambda_0/v_0} = 235.2 $.
		Parameters other than those in TABLE.\ref{table_fixed_parameters} are as follows:
		time increment $ h/\K{\Lambda_0/v_0} = 0.000555$,
		$ \lambda_0 = 6.0 $, $ R/\Lambda_0=6.3 $ and $ \xi/v_0=2.8\times10^{-2} $.
		The black line indicates the outer edge of the smoothed boundary.
		}
	\label{result_profile_omega}
\end{figure}

\begin{figure}[btp]
	\centering
	\includegraphics[width=0.50\textwidth]{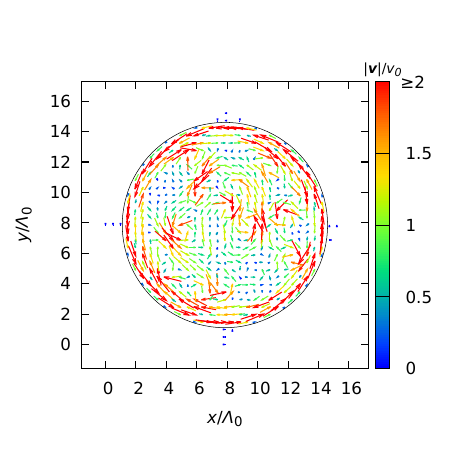}
	\caption{
		A typical simulation snapshot of the velocity field $ \bm{v}/v_0 $ at the time $ t/\K{\Lambda_0/v_0} = 235.2 $.
		Parameters other than those in TABLE.\ref{table_fixed_parameters} are the same as in FIG.\ref{result_profile_omega}.
		The black line indicates the outer edge of the smoothed boundary.
		Velocity arrows are drawn at intervals of 8 lattice points in each direction.
		}
	\label{result_profile_velocity}
\end{figure}

Using the simulation method explained in Sec.\ref{numerical_method}, we successfully realized the edge current. 
FIG.\ref{result_profile_omega} and \ref{result_profile_velocity} are typical snapshots of the vorticity and velocity field, respectively.
See ref.\cite{SupplementalMaterial} for the time evolution of the vorticity and velocity field.
We can observe
turbulence-like behavior in the bulk regions and uni-directional flow (edge current) at the boundary.
Focusing on the boundary region in FIG.\ref{result_profile_velocity}, we find the emergence of a
counter-rotating double layer which is reminiscent of the one reported in the experiment of Wioland et al. \cite{wioland2013confinement}.
The emergence of this counter-rotating double layer in our simulations can be explained as follows: The TTSH equation has a
characteristic length scale $\Lambda_0$ determined by its $\Gamma$-terms and
the direction of the velocity field switches over the distance $\sim\Lambda_0$.
Thus, if clockwise/counterclockwise edge current emerges at the boundary, respectively, counterclockwise/clockwise flow emerges inside, leaving the distance $\sim\Lambda_0$
from the outer counter-rotating layer.
The inner turbulent flow appears to destroy further counter-rotating layers that would be expected to emerge from the same argument,
leaving only the double layers in our simulations.

Now let us comment on the relevant length scales in the experiment of Wioland et al. \cite{wioland2013confinement}. The thickness of the boundary layer ($\sim 4\R{\mu m}$) has
almost the same value as the length of each single bacterium ($\sim 5\R{\mu m}$) and they argue that the bacteria trapped in the outer layer generate the backflow
and it stabilizes the opposite-directional bulk flow.
Here, recall that, as already mentioned in Sec.\ref{introduction}, the typical vortex size of bacterial turbulence (=the spacing between counter-rotating double layers in our simulations)
is larger than the length of single bacterium.
Therefore, the mechanisms of the emergence of the double layer in our simulations and the experiment by Wioland et al. are different.

To determine the direction of the edge current quantitatively, let us introduce the following quantity:
\begin{equation}
	\braket{v^{\R{tan}}}\equiv\F{1}{N_\R{b}}\sum_{i\in\R{boundary}}v^{\R{tan}}\K{i},
\end{equation}
where $ v^{\R{tan}}\K{=\K{\bm{n}\times\bm{v}}_z} $ is the tangential component of the velocity at the boundary,
$i$ is the lattice label,
the summation $\sum_{i\in\R{boundary}}$ is taken over the lattice points in the boundary region
defined by eq.(\ref{cutoff_phi}),
and $N_{\R{b}}$ is the total number of lattice points in the boundary region.
$\braket{v^{\R{tan}}}>0$ and $<0$ corresponds to the counterclockwise and clockwise edge current, respectively.

\begin{figure}[btp]
	\centering
	\includegraphics[width=0.5\textwidth]{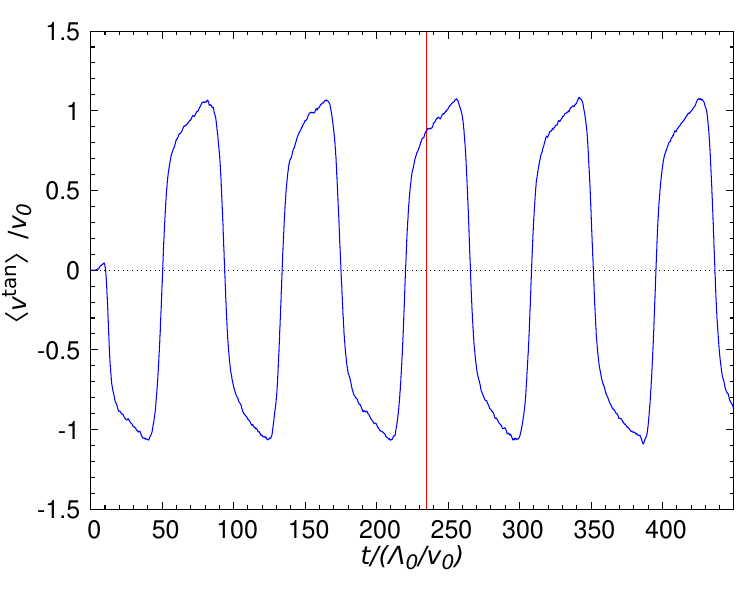}
	\caption{
		A typical example of the time evolution of $\braket{v^{\R{tan}}}$.
		Parameters other than those in TABLE.\ref{table_fixed_parameters} are the same as in FIG.\ref{result_profile_omega}.
		The vertical red line indicates the time at which
		the snapshots (FIG.\ref{result_profile_omega} and \ref{result_profile_velocity}) are obtained.
		}
		\label{v_tan_theta_average_tmev_R120}
\end{figure}

An example of the time evolution of $\braket{v^{\R{tan}}}$ is shown in FIG.\ref{v_tan_theta_average_tmev_R120}, which demonstrates
that the sign of $ \braket{v^{\R{tan}}} $ periodically changes.
This means that the direction of the edge current temporally oscillates.

\subsection{What causes the oscillation?}
\label{WhatCauseTheOscillation}
\begin{figure}[btp]
	\centering
	\includegraphics[width=0.50\textwidth]{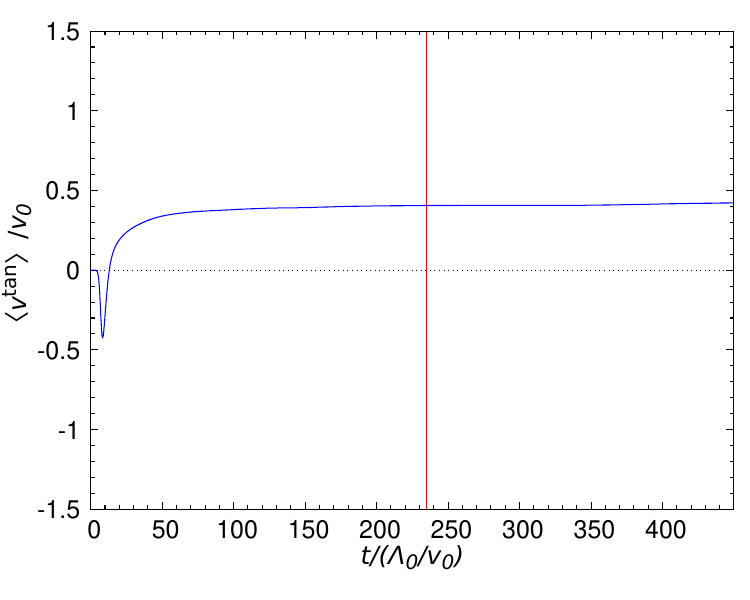}
	\caption{
		A typical example of the time evolution of $\braket{v^{\R{tan}}}$ for $ \lambda_0 =0$.
		Parameters other than those in TABLE.\ref{table_fixed_parameters} are the same as in FIG.\ref{result_profile_omega} except the value of $\lambda_0$.
		The vertical red line indicates the time at which the snapshots (FIG.\ref{vorticity_field_lambda_0_zero} and \ref{velocity_field_lambda_0_zero}) are obtained.
		}
			\label{v_tan_theta_average_lambda_0_zero}
\end{figure}

\begin{figure}[btp]
	\centering
	\includegraphics[width=0.50\textwidth]{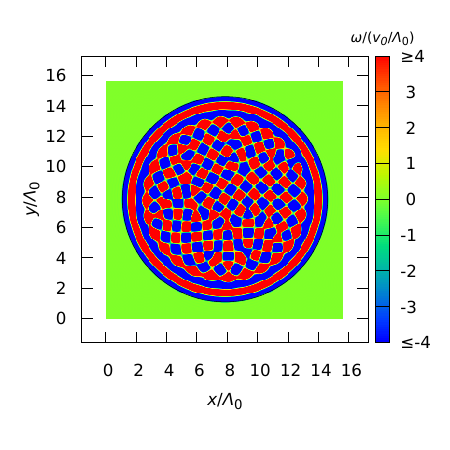}
	\caption{
		A typical simulation snapshot of the vorticity field
		$ \omega/\K{v_0/\Lambda_0}=\K{\bm{\nabla}\times\bm{v}}_z/\K{v_0/\Lambda_0} $ for $\lambda_0=0$
		at the time $ t/\K{\Lambda_0/v_0} = 235.2 $.
		Parameters other than those in TABLE.\ref{table_fixed_parameters} are the same as in FIG.\ref{result_profile_omega}
		except the value of $\lambda_0$.
		The black line indicates the outer edge of the smoothed boundary.
		In the bulk region, we can observe the stationary vortex lattice which has been observed in ref.\cite{Dunkel2013NewJPhys}.
		}
		\label{vorticity_field_lambda_0_zero}
\end{figure}

\begin{figure}[btp]
	\centering
	\includegraphics[width=0.50\textwidth]{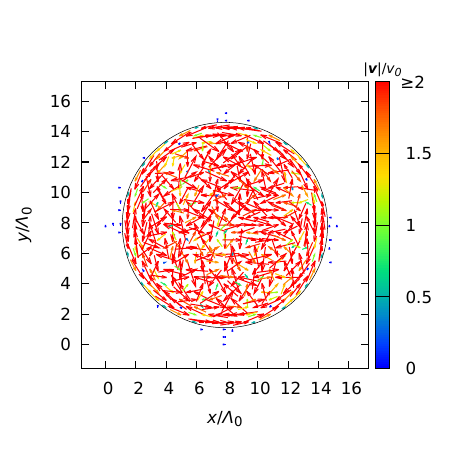}
	\caption{
		A typical simulation snapshot of the velocity field $ \bm{v}/v_0 $ for $\lambda_0=0$
		at the time $ t/\K{\Lambda_0/v_0} = 235.2 $.
		Parameters other than those in TABLE.\ref{table_fixed_parameters} are the same as in FIG.\ref{result_profile_omega}
		except the value of $\lambda_0$.
		The black line indicates the outer edge of the smoothed boundary.
		Velocity arrows are drawn at intervals of 8 lattice points in each direction.
		}
		\label{velocity_field_lambda_0_zero}
\end{figure}

Here, one simple question arises:
What causes the oscillation? To be more specific,
which term in the TTSH equation causes the oscillation?
To answer this question,
let us focus on the functional-derivative form of the TTSH equation (eq.(\ref{TTSH_eq_fluid})
with the functional $ \mathscr{F} $ (eq.(\ref{eq:free_energy_functional_slip}))):
\begin{equation}
	\label{TTSH_eq_fluid2}
	\partial_{t}\bm{v}=-\F{\delta\mathscr{F}}{\delta\bm{v}}-\lambda_{0}\bm{v}\cdot\bm{\nabla}\bm{v}.
\end{equation}
Without the $ \lambda_0 $-term, eq.(\ref{TTSH_eq_fluid2})
would become $ \partial_t\bm{v} = -\delta\mathscr{F}/\delta\bm{v} $ and
the velocity field would settle in a stationary state which minimizes the functional $ \mathscr{F} $.
Thus, oscillatory behavior is not expected to occur without the $ \lambda_0 $-term.
FIG.\ref{v_tan_theta_average_lambda_0_zero}, \ref{vorticity_field_lambda_0_zero} and \ref{velocity_field_lambda_0_zero}
show the results of simulation with $ \lambda_0=0 $ and we can confirm that oscillation does not occur.
Therefore, we should attribute the temporal oscillation to the $ \lambda_0 $-term.
The oscillation occurs for small but finite $\lambda_0$, and the discussion on the threshold value is given in Appendix \ref{threshold_lambda_0}.
Here, let us give several comments on the vorticity and velocity field for $\lambda_0=0$:
Looking at the vorticity field (FIG.\ref{vorticity_field_lambda_0_zero}),
we can observe the vortices whose linear size rescaled by $\Lambda_0$ is order $\sim 1$.
In the velocity field (FIG.\ref{velocity_field_lambda_0_zero}), the magnitude of the velocity rescaled by $v_0$ is order $\sim 1$.
Note that $\Lambda_0$ and $v_0$ are determined, respectively, by the $\Gamma$-terms and $\alpha,\beta$-term in $\mathscr{F}$.
From the above discussions, we can confirm that, without the $\lambda_0$-term, the result profiles become ones which minimize the functional $\mathscr{F}$.

\begin{figure}[btp]
	\centering
	\includegraphics[width=0.20\textwidth]{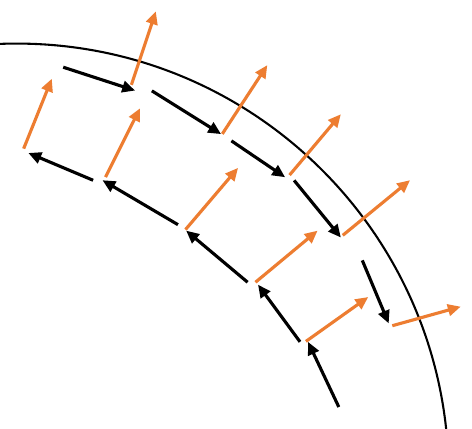}
	\caption{
		A schematic of the velocity $\bm{v}$ (black arrow) and
		the adevection term $-\lambda_0\bm{v}\cdot\bm{\nabla}\bm{v}$ (orange arrow) at the boundary.
		The curved black line indicates the outer edge of the smoothed boundary.
		}
		\label{advection_at_boundary}
\end{figure}

Next we give a simple argument on how the $\lambda_0$-term causes the oscillation:
The velocity field at the boundary and the inner counter-rotating layer can be
approximately written by $\bm{v}\sim c\K{r}\bm{e}_\theta$, where $\bm{e}_\theta$ is 
the azimuthal unit vector of
two-dimensional polar coordinate whose origin is at the center of the circular domain.
Using this $\bm{v}$, we can calculate the advection term $-\lambda_0\bm{v}\cdot\bm{\nabla}\bm{v}$ and obtain
\begin{equation}
	\label{advection_term}
	-\lambda_0\bm{v}\cdot\bm{\nabla}\bm{v}\sim\F{\lambda_0 \CK{c\K{r}}^2}{R}\bm{e}_r,
\end{equation}
where $\bm{e}_{r}$ is the radial unit vector of the two-dimensional polar coordinate.
Therefore, the advection term has outward radial profile regardless of the sign of $c\K{r}$
(i.e. the direction of the flow) (see FIG.\ref{advection_at_boundary}) and this term will exert the velocity field
to rotate toward the opposite direction.
Here, note that the outer flow never has radial component because of the boundary condition and then it cannot rotate.
However, the inner counter-rotating flow can have radial component and then it can rotate and switch its direction.

\subsection{Oscillation frequency vs. $ R $, $ \xi $ and $ \lambda_0 $}
\label{omega_c_vs_parameters}
Next, let us investigate the relations between the behavior of the edge current oscillation and the parameters $R$, $\xi$ and $\lambda_0$ with particular focus on the oscillation frequency.
To characterize the oscillation frequency, let us introduce the characteristic angular frequency $\omega_{\R{c}}$ defined by
\begin{equation}
	\label{center_of_power_spectrum}
	\omega_{\R{c}}\equiv\F{\sum_n\ABS{\R{FT}\BK{\braket{v^{\R{tan}}}}_n}^2 \omega_n}{\sum_n\ABS{\R{FT}\BK{\braket{v^{\R{tan}}}}_n}^2},
\end{equation}
where $ \R{FT}\BK{\braket{v^{\R{tan}}}}_n $ is the $ n $-th component of the Fourier transform of $ \braket{v^{\R{tan}}} $
and $ \omega_n $ is the angular frequency of the $ n $-th Fourier mode.

\begin{figure*}[tp]
	\begin{center}
		\includegraphics[width=0.8\textwidth]{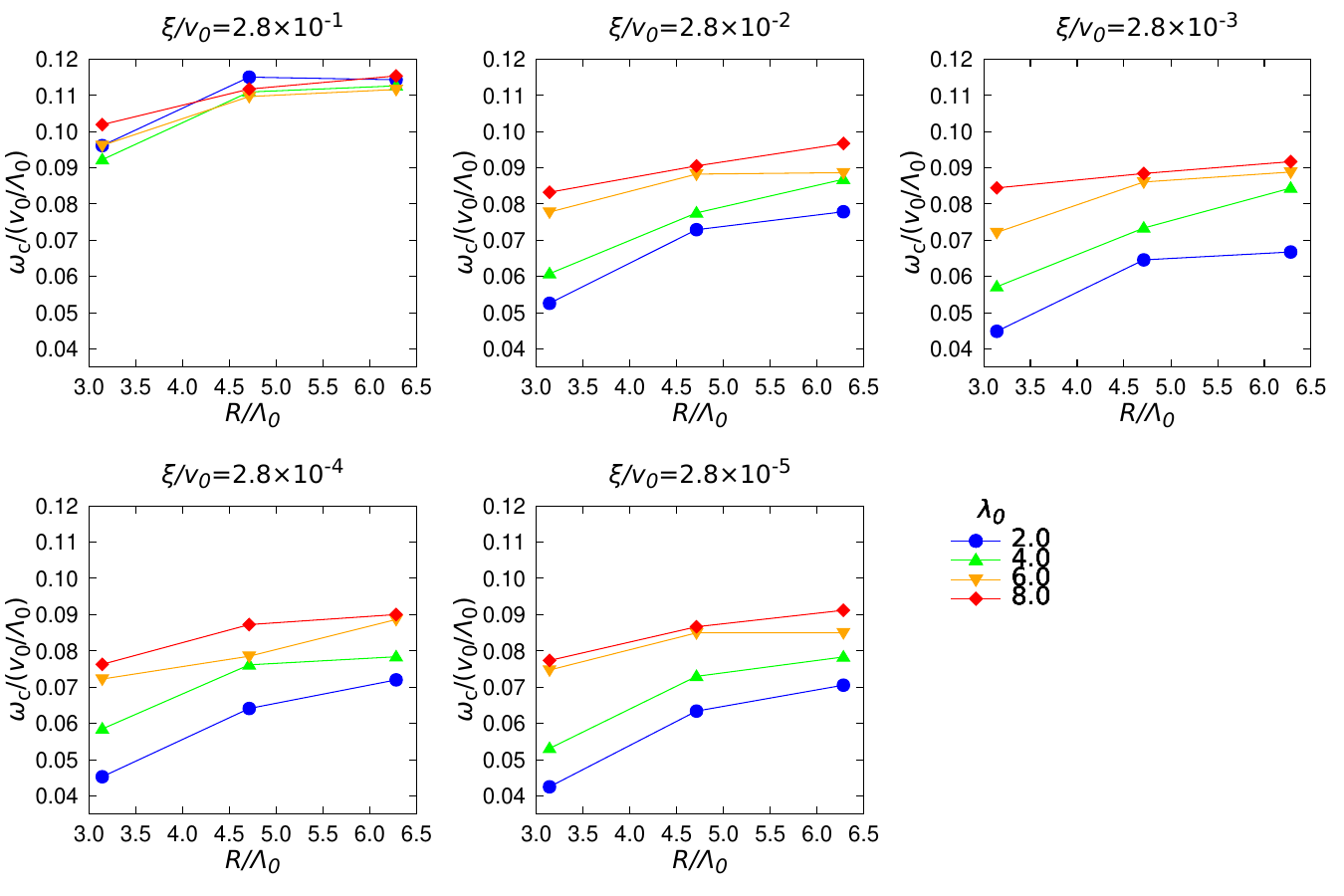}
		\caption{
			The radius of the fluid region $R$ versus the characteristic angular frequency $\omega_{\R{c}}$ for G1 geometry.
			All simulations are performed with $h/\K{\Lambda_0/v_0}=0.000555$ and parameters listed in TABLE.\ref{table_fixed_parameters}.
			}
		\label{xaxis_R_varied_lambda_0_multiplot_xi}
	\end{center}
\end{figure*}

\begin{figure*}[bp]
	\begin{center}
		\includegraphics[width=0.533\textwidth]{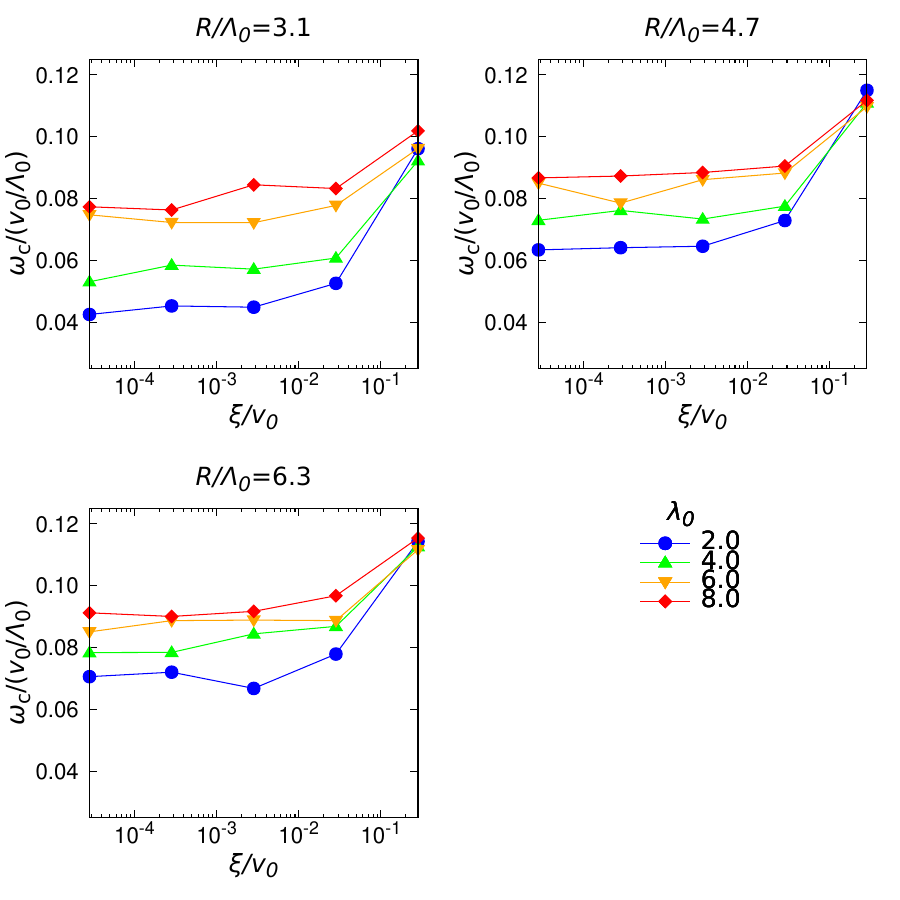}
		\caption{
			The drag coefficient $\xi$ versus the characteristic angular frequency $\omega_{\R{c}}$ for G1 geometry.
			All simulations are performed with $h/\K{\Lambda_0/v_0}=0.000555$ and parameters listed in TABLE.\ref{table_fixed_parameters}.
			}
		\label{xaxis_xi_varied_lambda_0_multiplot_R}
	\end{center}
\end{figure*}

We performed simulations with all combinations of the following $R$, $\xi$ and $\lambda_0$:
$R/\Lambda_0=1.0, 1.6, 3.1, 4.7, 6.3$, $\xi/v_0=2.8\times10^{-1}, 2.8\times10^{-2}, 2.8\times10^{-3}, 2.8\times10^{-4}, 2.8\times10^{-5}$
and $\lambda_0=0, 2.0, 4.0, 6.0, 8.0$.
The results are shown in FIG.\ref{xaxis_R_varied_lambda_0_multiplot_xi} and \ref{xaxis_xi_varied_lambda_0_multiplot_R}.
When $\lambda_0=0$ or $R/\Lambda_0=1.0, 1.6$, the oscillation of $\braket{v^{\R{tan}}}$ does not occur as in FIG.\ref{v_tan_theta_average_lambda_0_zero}
and then we do not plot them in FIG.\ref{xaxis_R_varied_lambda_0_multiplot_xi} and \ref{xaxis_xi_varied_lambda_0_multiplot_R}.
We can observe that $\omega_{\R{c}}$ tends to increase with increase of $\lambda_0$, $R$ and $\xi$.
Here, let us comment on the dependence on $\xi$.
$\omega_{\R{c}}$ tends to increase with $\xi$, but is almost constant in the small-$\xi$ region.
This is probably because, at $\xi/v_0 \sim 10^{-2}$,
the boundary condition has already reduced to the free-slip one
(Recall the discussion presented in Sec.\ref{model}: In the limit $\xi\rightarrow 0$, the boundary condition will
reduce to the stress-free slip condition).

Here, note that the larger the advection strength $\lambda_0$ is,
the more excited the turbulent behavior in the bulk become.
Furthermore, the larger the drag coefficient $\xi$ is, the more subject to the bulk turbulence the edge current is.
From the above arguments, we can say that the edge current oscillation will be related to the turbulent behavior
in the bulk region.
The relations between $\omega_{\R{c}}$ and $R$ will be discussed
after the simulations in two more geometries in Sec.\ref{other_geometries}.


\subsection{Other geometries}
\label{other_geometries}
\begin{figure}[btp]
	\centering
	\includegraphics[width=0.5\textwidth]{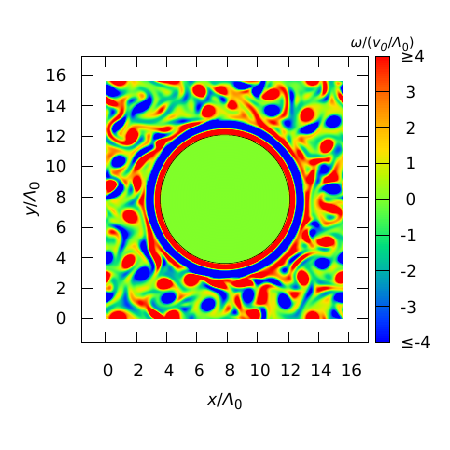}
	\caption{
		A typical simulation snapshot of the vorticity field $ \omega/\K{v_0/\Lambda_0}=\K{\bm{\nabla}\times\bm{v}}_z/\K{v_0/\Lambda_0} $
		at the time $ t/\K{\Lambda_0/v_0} = 312.9 $ for G2 geometry.
		Parameters other than those in TABLE.\ref{table_fixed_parameters} are as follows:
		time increment $ h/\K{\Lambda_0/v_0} = 0.000370$,
		$ \lambda_0 = 4.0 $, $ R/\Lambda_0=4.7 $ and $ \xi/v_0=2.8\times10^{-2} $.
		The black line indicates the outer edge of the smoothed boundary.
		}
		\label{opposite_curvature_vorticity}
\end{figure}

\begin{figure}[btp]
	\centering
	\includegraphics[width=0.5\textwidth]{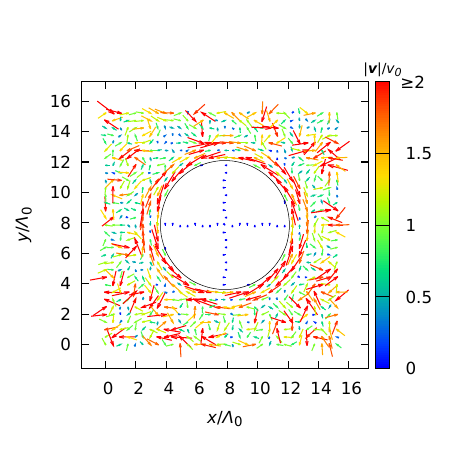}
	\caption{
		A typical simulation snapshot of the velocity field $ \bm{v}/v_0 $ at the time $ t/\K{\Lambda_0/v_0} = 312.9 $ for G2 geometry.
		Parameters other than those in TABLE.\ref{table_fixed_parameters} are the same as in FIG.\ref{opposite_curvature_vorticity}.
		The black line indicates the outer edge of the smoothed boundary.
		Velocity arrows are drawn at intervals of 8 lattice points in each direction.
		}
		\label{opposite_curvature_velocity}
\end{figure}

\begin{figure}[btp]
	\centering
	\includegraphics[width=0.5\textwidth]{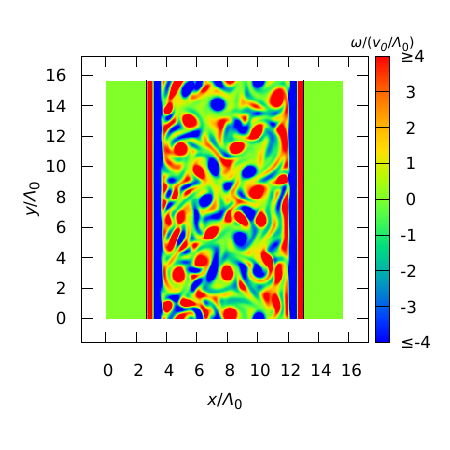}
	\caption{
		A typical simulation snapshot of the vorticity field $ \omega/\K{v_0/\Lambda_0}=\K{\bm{\nabla}\times\bm{v}}_z/\K{v_0/\Lambda_0} $
		at the time $ t/\K{\Lambda_0/v_0} = 274.6 $ for G3 geometry.
		Parameters other than those in TABLE.\ref{table_fixed_parameters} are as follows:
		time increment $ h/\K{\Lambda_0/v_0} = 0.000370$,
		$ \lambda_0 = 4.0 $, $ R/\Lambda_0=4.7 $ and $ \xi/v_0=2.8\times10^{-2} $.
		Note that the spacing of two walls is $2R/\Lambda_0$.
		The black line indicates the outer edge of the smoothed boundary.
		}
		\label{zero_curvature_vorticity}
\end{figure}

\begin{figure}[btp]
	\centering
	\includegraphics[width=0.5\textwidth]{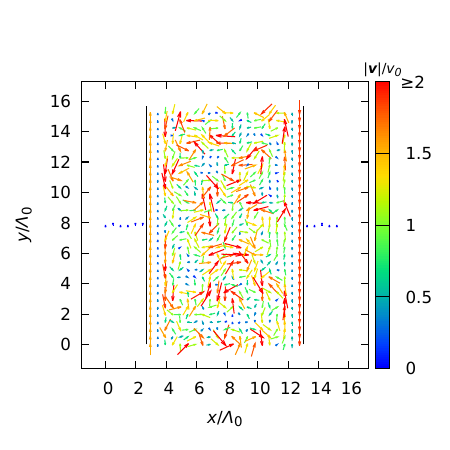}
	\caption{
		A typical simulation snapshot of the velocity field $ \bm{v}/v_0 $ at the time $ t/\K{\Lambda_0/v_0} = 274.6 $ for G3 geometry.
		Parameters other than those in TABLE.\ref{table_fixed_parameters} are the same as in FIG.\ref{zero_curvature_vorticity}.
		The black line indicates the outer edge of the smoothed boundary.
		Velocity arrows are drawn at intervals of 8 lattice points in each direction.
		}
		\label{zero_curvature_velocity}
\end{figure}

The next question is whether the edge current oscillation occurs in other geometries.
To answer this question, we have performed the simulations in two more geometries.
Let us call the geometry used in FIG.\ref{result_profile_omega}-\ref{velocity_field_lambda_0_zero} ``G1''.
The first one of the two new geometries (hereinafter called ``G2'', see FIG.\ref{opposite_curvature_vorticity}
and \ref{opposite_curvature_velocity}) is the case where the solid and fluid regions are swapped
in G1.
The second one (hereinafter called ``G3'', see FIG.\ref{zero_curvature_vorticity}
and \ref{zero_curvature_velocity}) 
is the case where the active fluid is sandwiched by two parallel straight walls perpendicular to the $x$-axis and periodic boundary condition is imposed in the $y$-direction.

We can implement the above two new geometries, G2 and G3, only by replacing the smoothed profile $\phi$ (eq.(\ref{phi_tanh})) of G1.
This is an advantage of the SPM.

The smoothed profile for G2, $\tilde{\phi}$, is
\begin{equation}
	\label{phi_opposite_curvature}
	\tilde{\phi} = 1-\phi = \F{1}{2} - \F{1}{2}\tanh\F{R-r}{d},
\end{equation}
where $R$ is the radius of the circular solid domain and $r$, $d$ are the same
ones already used in eq.(\ref{phi_tanh}).
The smoothed profile for G3, $\tilde{\tilde{\phi}}$, can be made by replacing $r$ in $\phi$ by $x$:
\begin{equation}
	\label{phi_straight_wall}
	\tilde{\tilde{\phi}} = \F{1}{2}\tanh\F{R-x}{d}+\F{1}{2},
\end{equation}
where $2R$ is the spacing of the two walls
and $d$ is the same as in eq.(\ref{phi_tanh}).

\begin{figure}[btp]
	\centering
	\includegraphics[width=0.50\textwidth]{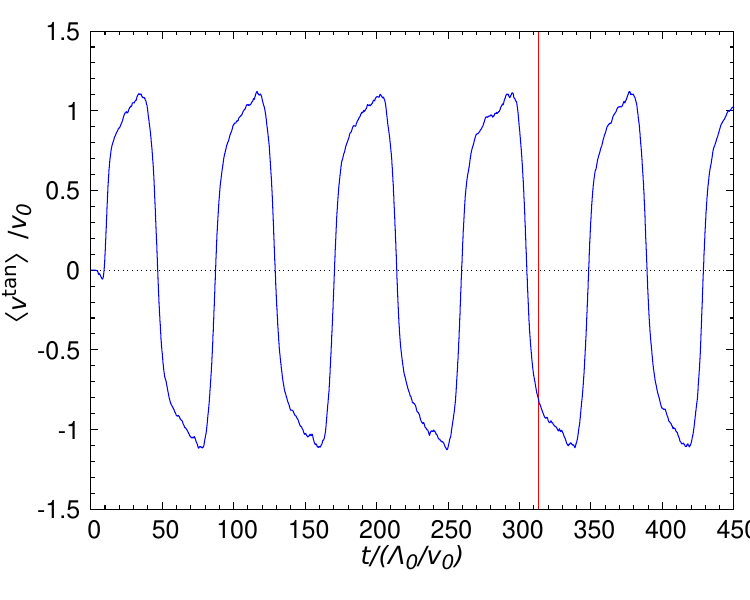}
	\caption{
		A typical example of the time evolution of $ \braket{v^{\R{tan}}} $ for G2 geometry.
		Parameters other than those in TABLE.\ref{table_fixed_parameters} are the same as in FIG.\ref{opposite_curvature_vorticity}.
		The vertical red line indicates 
		the time at which the snapshots (FIG.\ref{opposite_curvature_vorticity} and \ref{opposite_curvature_velocity}) are obtained.
		}
		\label{opposite_curvature_tmev_of_v_tan}
\end{figure}

\begin{figure}[btp]
	\centering
	\includegraphics[width=0.5\textwidth]{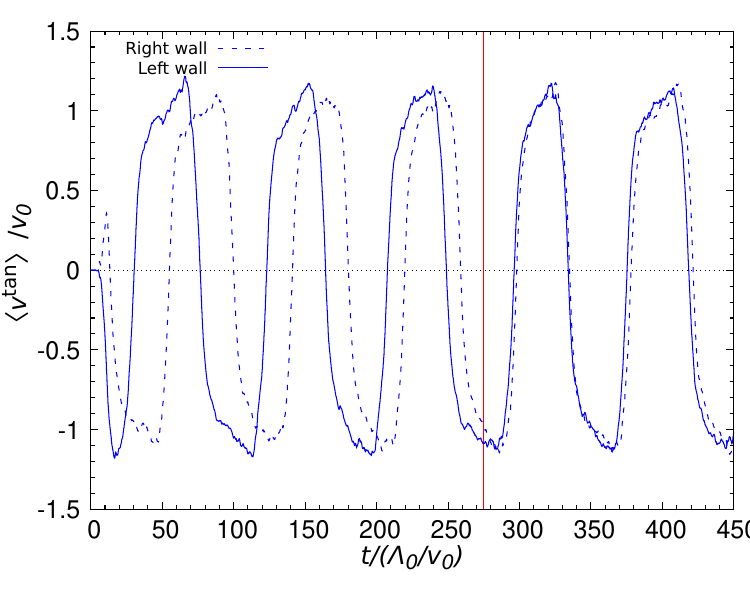}
	\caption{
		A typical example of the time evolution of $\braket{v^{\R{tan}}}$ for G3 geometry.
		Parameters other than those in TABLE.\ref{table_fixed_parameters} are the same as in FIG.\ref{zero_curvature_vorticity}.
		The vertical red line indicates the time
		at which the snapshots (FIG.\ref{zero_curvature_vorticity} and \ref{zero_curvature_velocity}) are obtained.
		}
		\label{zero_curvature_tmev_of_v_tan}
\end{figure}

In FIG.\ref{opposite_curvature_vorticity}, \ref{opposite_curvature_velocity},
\ref{zero_curvature_vorticity} and \ref{zero_curvature_velocity},
we can observe that the edge current emerges again in G2 and G3.
Furthermore, FIG.\ref{opposite_curvature_tmev_of_v_tan} and \ref{zero_curvature_tmev_of_v_tan} show
that the edge current oscillation occurs also in G2 and G3.
See ref.\cite{SupplementalMaterial} for the time evolution of the vorticity and velocity field.
G3 geometry is also the subject of previous studies \cite{duclos2014perfect,li2017mechanism}, and their results are compared with ours in Appendix \ref{behavior_in_narrow_channel}.
Here, let us comment on what causes the oscillation. In the case of G2, we can apply the same argument as in Sec.\ref{WhatCauseTheOscillation}.
In the geometry G3, applying the similar argument as in Sec.\ref{WhatCauseTheOscillation}, we obtain the result $-\lambda_0\bm{v}\cdot\bm{\nabla}\bm{v}\sim 0$ as follows:
The velocity at the boundary and inner counter-rotating layer can be approximately written by $\bm{v}\sim c\K{x}\bm{e}_y$
and $-\lambda_0\bm{v}\cdot\bm{\nabla}\bm{v}\sim-\lambda_0 c\K{x}\partial/\partial y\K{c\K{x}\bm{e}_y}=0$.
However, in the inner counter-rotating layer, the velocity is affected by the inner turbulent flow and distorted. Therefore, the inner layer locally has finite curvature and then
we can apply the same argument as in Sec.\ref{WhatCauseTheOscillation} locally.

\begin{figure*}[tp]
	\begin{center}
		\includegraphics[width=0.8\textwidth]{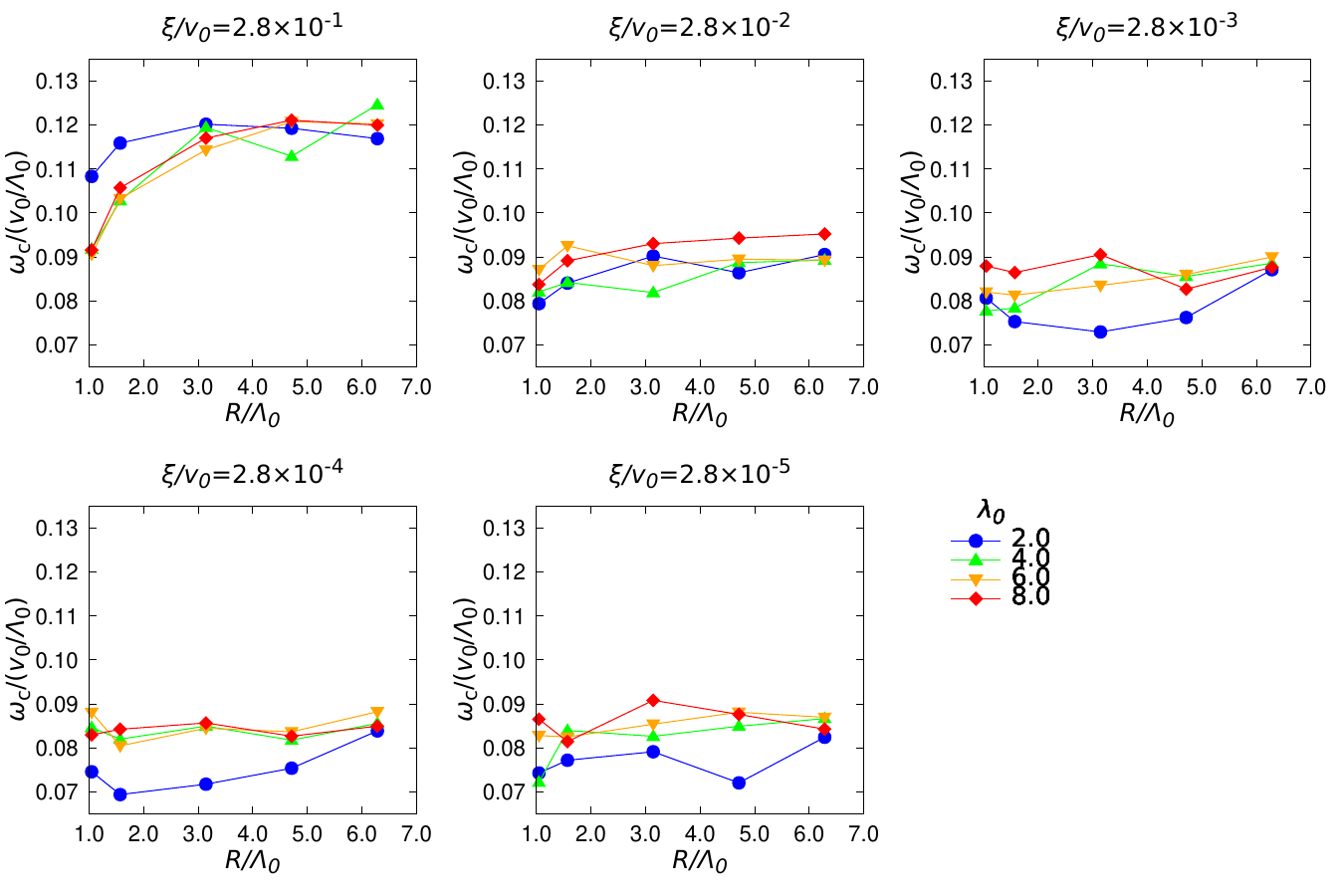}
		\caption{
			The radius of the fluid region $R$ versus the characteristic angular frequency $\omega_{\R{c}}$ for G2 geometry.
			All simulations are performed with $h/\K{\Lambda_0/v_0}=0.000370$ and parameters listed in TABLE.\ref{table_fixed_parameters}.
			}
		\label{xaxis_R_varied_lambda_0_multiplot_xi_G2}
	\end{center}
\end{figure*}

\begin{figure*}[bp]
	\begin{center}
		\includegraphics[width=0.8\textwidth]{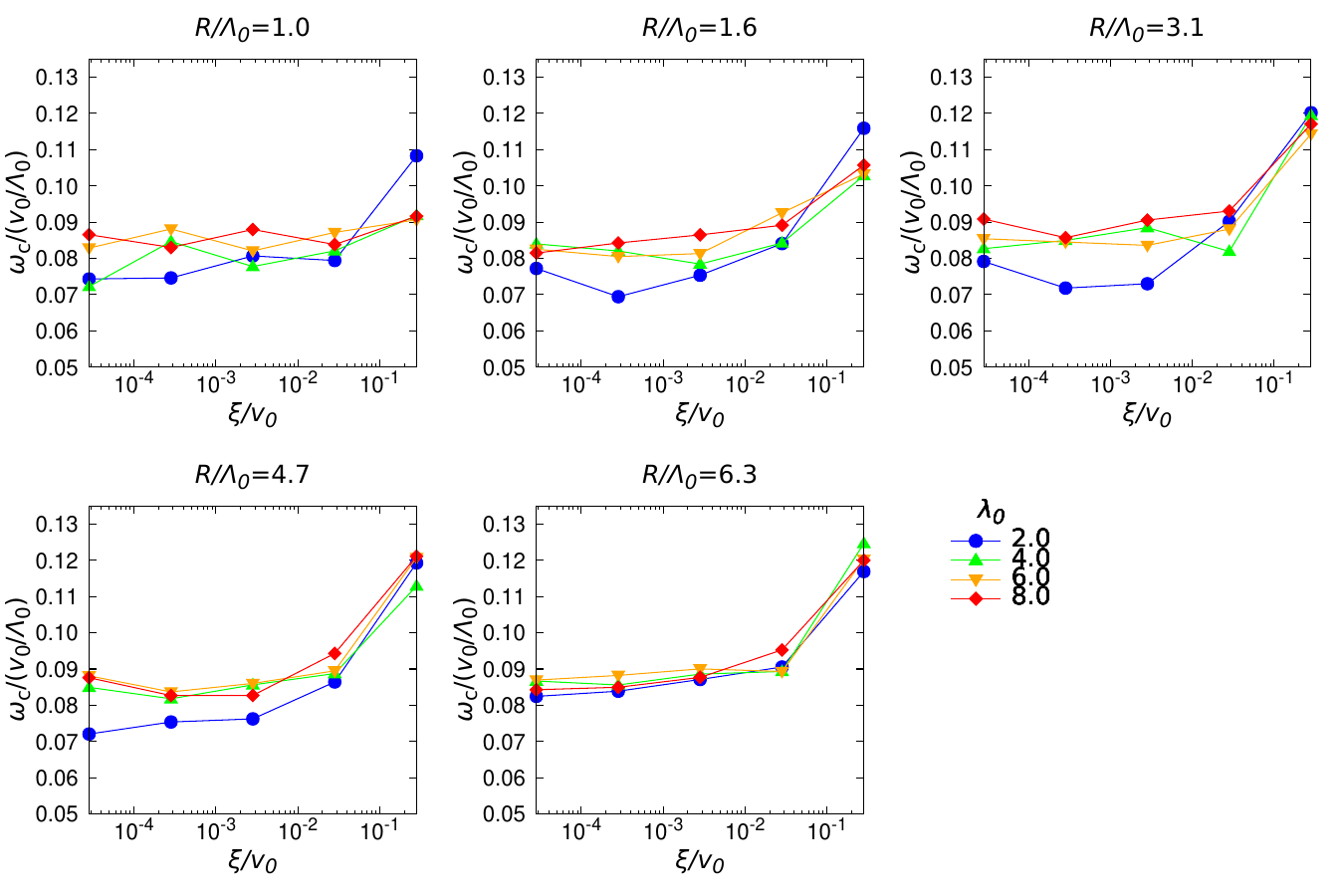}
		\caption{
			The drag coefficient $\xi$ versus the characteristic angular frequency $\omega_{\R{c}}$ for G2 geometry.
			All simulations are performed with $h/\K{\Lambda_0/v_0}=0.000370$ and parameters listed in TABLE.\ref{table_fixed_parameters}.
			}
		\label{xaxis_xi_varied_lambda_0_multiplot_R_G2}
	\end{center}
\end{figure*}

\begin{figure*}[tp]
	\begin{center}
		\includegraphics[width=0.8\textwidth]{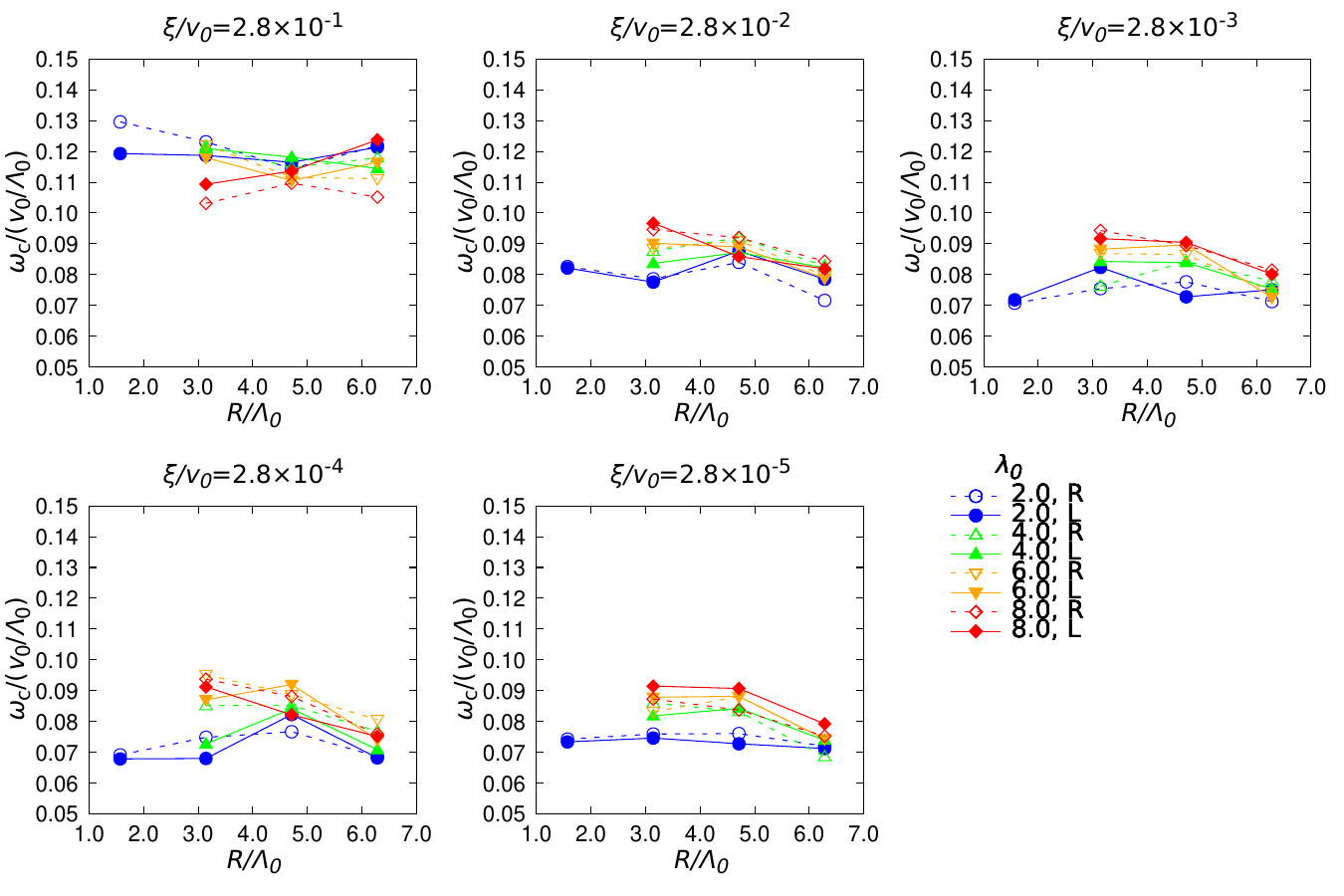}
		\caption{
			Half of the spacing of two walls $R$ 
			versus the characteristic angular frequency $\omega_{\R{c}}$ for G3 geometry.
			$\omega_{\R{c}}$ is calculated in each (right and left) wall.
			``R'' and ``L'' in the legend indicates the right and left wall, respectively.
			All simulations are performed with $h/\K{\Lambda_0/v_0}=0.000370$ and parameters listed in TABLE.\ref{table_fixed_parameters}.
			}
			\label{xaxis_R_varied_lambda_0_multiplot_xi_G3}
	\end{center}
\end{figure*}

\begin{figure*}[bp]
	\begin{center}
		\includegraphics[width=0.8\textwidth]{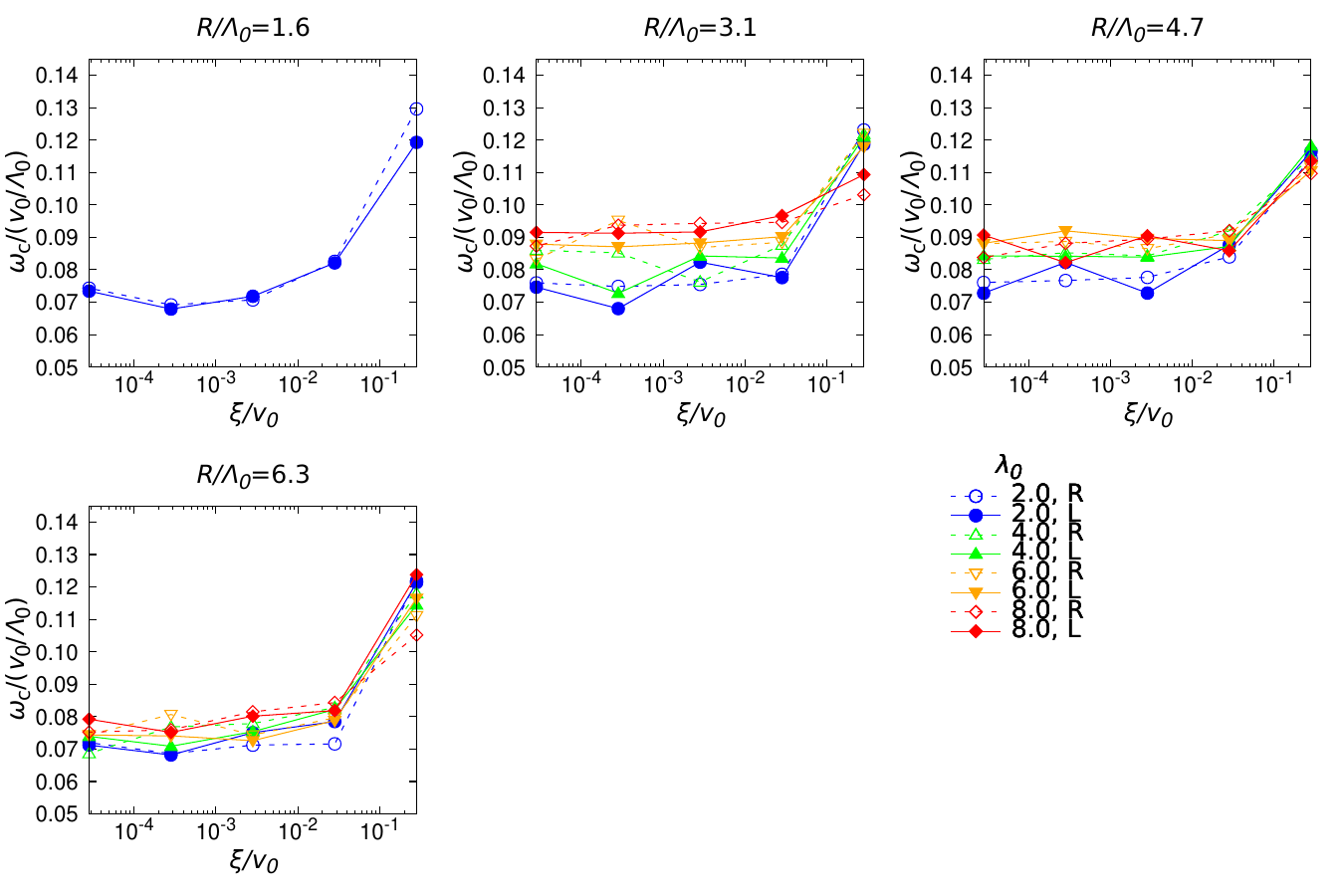}
		\caption{
			The drag coefficient $\xi$
			versus the characteristic angular frequency $\omega_{\R{c}}$ for G3 geometry.
			$\omega_{\R{c}}$ is calculated in each (right and left) wall.
			``R'' and ``L'' in the legend indicates the right and left wall, respectively.
			All simulations are performed with $h/\K{\Lambda_0/v_0}=0.000370$ and parameters listed in TABLE.\ref{table_fixed_parameters}.
			}
			\label{xaxis_xi_varied_lambda_0_multiplot_R_G3}
	\end{center}
\end{figure*}

Next, we calculate the characteristic angular frequency $\omega_\R{c}$
and investigate the relation between $\omega_{\R{c}}$ and $R$, $\xi$, $\lambda_0$
\MYBLACK{
	for G2 (FIG.\ref{xaxis_R_varied_lambda_0_multiplot_xi_G2} and \ref{xaxis_xi_varied_lambda_0_multiplot_R_G2})
	and G3 (FIG.\ref{xaxis_R_varied_lambda_0_multiplot_xi_G3} and \ref{xaxis_xi_varied_lambda_0_multiplot_R_G3}).
}
The values of $R$, $\xi$ and $\lambda_0$ are the same as in G1.
In both geometries, as expected, for $\lambda_0=0$ edge current oscillation does not occur.
Unlike in the case of G1, in G2 geometry,
oscillations occur for all $R$'s.
\MYBLACK{In} G3 geometry,
for $R/\Lambda_0=1.0$ oscillation \MYBLACK{does} not occur,
for $R/\Lambda_0=1.6$ oscillation is observed only when $\lambda_0=2.0$
and for other $R$'s oscillation \MYBLACK{occurs} for all values of $\lambda_0$.
As in the case of G1, we plot $\omega_\R{c}$ only for the cases where oscillation is observed.
\MYBLACK{
	The dependence of $\omega_{\R{c}}$ on $\xi$ for G2 and G3 is similar to \MYBLACK{that} for G1
}
(see FIG.\ref{xaxis_xi_varied_lambda_0_multiplot_R_G2} and \ref{xaxis_xi_varied_lambda_0_multiplot_R_G3}).
However, the dependence on $R$ and $\lambda_0$ looks different from the one \MYBLACK{for} G1 geometry.
First, let us comment on the relations between $\omega_{\R{c}}$ and $\lambda_0$.
In G3 geometry (FIG.\ref{xaxis_R_varied_lambda_0_multiplot_xi_G3}
or \ref{xaxis_xi_varied_lambda_0_multiplot_R_G3}),
$\omega_{\R{c}}$ tends to \MYBLACK{increase slightly} with \MYBLACK{the increase of} $\lambda_0$
but the tendency is not so clear as in G1 geometry.
In G2 geometry (FIG.\ref{xaxis_R_varied_lambda_0_multiplot_xi_G2}
or \ref{xaxis_xi_varied_lambda_0_multiplot_R_G2}),
$\omega_{\R{c}}$ \MYBLACK{is insensitive to the variation of} $\lambda_0$.
From the relations between $\omega_{\R{c}}$ and $\lambda_0$ in \MYBLACK{the} three geometries,
we can say that \MYBLACK{the more strongly confined the active fluids are},
the greater effect the $\lambda_0$-term has on the edge current oscillation.
Next, let us mention the relations between $\omega_{\R{c}}$ and $R$.
In G2 geometry (FIG.\ref{xaxis_R_varied_lambda_0_multiplot_xi_G2}),
$\omega_{\R{c}}$ is almost constant.
In G3 geometry (FIG.\ref{xaxis_R_varied_lambda_0_multiplot_xi_G3}),
$\omega_{\R{c}}$ tends to \MYBLACK{decrease slightly} with $R$.
The typical value of $\omega_{\R{c}}$ for G3
is $\sim 0.08v_0/\Lambda_0$ (FIG.\ref{xaxis_R_varied_lambda_0_multiplot_xi_G3}), close to the
asymptotic value for G1 with large $R$.
This is natural from the fact that the $R\rightarrow\infty$ limit of the G1 geometry can be
regarded as the flat geometry G3.
Furthermore, the typical value of $\omega_{\R{c}}$ for G2 is close to the one for G3.
Those behavior of $\omega_{\R{c}}$ in response to the variation of $R$ suggests that the curvature of the
boundary, when positive and small enough (large-$R$ region in G1) or negative (G2),
does not play an important role in the oscillation of the edge current.
Finally, $\omega_{\R{c}}$'s tendency in G3 to slightly decrease with the increasing of $R$ can be
explained as follows:
In G3 geometry, the smaller $R$ is, the more narrow region the active fluids are confined in,
and then the more excited the turbulent behavior becomes.
Therefore $\omega_{\R{c}}$ become larger for smaller $R$.

\section{Conclusion}
\label{conclusion}
The previous works on TTSH simulations seem to be insufficient in that
although the importance of edge current in bacterial dynamics has been
reported in several experimental studies, the previous TTSH
simulations adopted a non-slip boundary condition and did not describe such bacterial motions.

In this paper, we focused on the bacterial behavior at the fluid-solid boundary and
adopted a slip boundary condition to investigate the effect of the boundary condition
on the bacterial dynamics.
To implement the slip boundary condition,
we proposed an extended TTSH model,
where a surface term which can be regarded as a energetic penalty
for the tangential component of velocity
is added to the functional in the TTSH equation.
Furthermore we applied the smoothed profile method to our model and performed
numerical simulations in three boundary geometries.
Our extended TTSH model successfully realized the edge current in three different boundary geometries.
Furthermore, we unexpectedly discovered the temporal oscillation of the direction of the edge current.
The edge current oscillation is observed in all of three boundary geometries.
By a simple argument based on the functional-derivative form of the TTSH equation,
the origin of the oscillation was identified as the $\lambda_0$-term
(advection term) in the TTSH equation.
Note that our argument identifies only the ``cause'' of the oscillation and its detailed ``mechanism'' is still an open question.

To determine the direction of the edge current, we introduced $\braket{v^{\R{tan}}}$,
the average of the tangential component of the velocity in boundary regions, and we traced its time evolution.
To characterize the oscillation of $\braket{v^{\R{tan}}}$ quantitatively,
we calculate the power spectrum of the time evolution of $\braket{v^{\R{tan}}}$
and then the characteristic angular frequency $\omega_{\R{c}}$.
We investigated the dependence of $\omega_{\R{c}}$ on three parameters
$R$, $\xi$ and $\lambda_0$.

We revealed the behavior of bacterial turbulence in contact with a slip boundary and
our work indicates that the boundary condition could play an important role in bacterial dynamics.

To our knowledge, oscillating edge current of active fluids has never been observed.
In our simulations only the velocity field in two dimensions is considered, and
the factors not taken into account in our study, say, the third dimension,
the spatial variation of the density, or other degrees of freedom, might suppress the oscillation.
How the oscillation could be promoted or suppressed could be an interesting direction of future studies.
We also hope that the oscillating edge current may be observed in a carefully performed future experiment.
\section*{Acknowledgement}
\label{acknowledgement}
The authors thank Prof. Yusuke T. Maeda and Dr. Kazusa Beppu for fluitfull discussions.
H.M. is supported by JST SPRING, Grant Number JPMJSP2136.
J.F. is supported by JSPS KAKENHI (Grants No. JP21H01049).
Substantial part of the computation
has been done using the facilities of the Supercomputer Center,
the institute for Solid State Physics,
the University of Tokyo.
This work was supported by the JSPS Core-to-Core Program ``Advanced
core-to-core network for the physics of self-organizing active matter
(JPJSCCA20230002)''.
\begin{appendices}
\section{Details of the calculation in sec.\ref{limit_d_zero}}
\label{appendix}
This appendix is devoted to proving that in the limit $d\rightarrow 0$, our extended TTSH equation
(eq.(\ref{TTSH_eq_fluid_phi}) with $ \mathscr{F}_{\phi}\BK{\bm v} $ given by (\ref{eq:free_energy_phi}))
reduces to eq.(\ref{TTSH_eq_fluid1}) in the fluid region,
eq.(\ref{equation_at_boundary}) at the fluid-solid boundary and eq.(\ref{equation_in_solid}) in the solid region.

Substituting eq.(\ref{eq:free_energy_phi}) into eq.(\ref{TTSH_eq_fluid_phi}) and executing the functional derivative, we obtain
\begin{align}
	\label{TTSH_eq_phi_2}
	\begin{split}
		\partial_t {\bm v} = &- \K{\bm{\nabla}\phi}q -\phi\bm{\nabla}q -\alpha\phi\bm{v} -\beta\phi\ABS{\bm v}^2 \bm{v}\\
		&+ \Gamma_0 \K{\bm{\nabla}\phi}\cdot\bm{\nabla}\bm{v}
		+ \Gamma_0 \phi\bm{\nabla}^2 \bm{v}\\
		&-\Gamma_2 \K{\bm{\nabla}^2 \phi}\bm{\nabla}^2 \bm{v}\\
		&- 2 \Gamma_2 \K{\bm{\nabla} \phi}\cdot\bm{\nabla} \bm{\nabla}^2 \bm{v}\\
		&- \Gamma_2 \phi \K{\bm{\nabla}^2}^2 \bm{v}\\
		&-\xi \ABS{\bm{\nabla}\phi}^{-1}\K{\K{\bm{\nabla}\phi}\cdot\Tensor{\epsilon}}\K{\K{\bm{\nabla}\phi}\times\bm{v}}_{z}\\
		&- \lambda_0 \phi \bm{v}\cdot\bm{\nabla}\bm{v}
	\end{split}
\end{align}
where $\Tensor{\epsilon}$ ($\K{\Tensor{\epsilon}}_{ij}=\epsilon_{ij}$)
is the two-dimentional Levi=Civita symbol,
whose definition has already given in Sec.\ref{practical_techniques}.

As discussed in \cite{Kanke_Sasaki_2013}, in the limit $d\rightarrow 0$,
\begin{align}
	\label{phi_step}
	\phi\K{\bm r} \rightarrow \Theta\K{R-r}= \left\{
		\begin{array}{ll}
			1 & \quad\text{for}\quad r<R \\
			0 & \quad\text{for}\quad r>R
		\end{array}
		\right.
		,
\end{align}
\begin{align}
	\label{nabla_phi}
	\bm{\nabla} \phi \rightarrow - \delta\K{r-R} {\bm n},
\end{align}
\begin{align}
	\label{abs_nabla_phi}
	\ABS{\bm{\nabla}\phi}\rightarrow\delta\K{r-R},
\end{align}
\begin{align}
	\label{normal_vector}
	\F{\bm{\nabla} \phi}{\ABS{\bm{\nabla} \phi}} \rightarrow - {\bm n}.
\end{align}
Hence, in the limit $d\rightarrow 0$, eq.(\ref{TTSH_eq_phi_2}) becomes
\begin{align}
	\begin{split}
		\label{TTSH_eq_phi_3}
		\partial_t {\bm v} = &\Theta\K{R-r}\CK{
			-\bm{\nabla}q -\K{\alpha + \beta \ABS{\bm v}^2}\bm{v} \right. \\
			& \left. \quad\quad\quad\quad +\Gamma_0 \bm{\nabla}^2 \bm{v} - \Gamma_2 \K{\bm{\nabla}^2}^2 \bm{v}\right. \\
			& \left. \quad\quad\quad\quad - \lambda_0 \bm{v}\cdot\bm{\nabla}\bm{v}
			}\\
			&+\delta\K{r-R}\CK{
				{\bm n}q - \Gamma_0 {\bm n}\cdot\bm{\nabla} \bm{v}\right. \\ 
				& \left. \quad\quad\quad\quad\quad+ 2 \Gamma_2 \bm{n} \cdot \bm{\nabla} \bm{\nabla}^2 {\bm v}\right. \\
				& \left. \quad\quad\quad\quad\quad-\xi \K{\bm{n}\cdot\Tensor{\epsilon}}\K{\bm{n}\times\bm{v}}_{z}}\\
				&+\Gamma_2 \CK{\bm{\nabla}\cdot\K{\delta\K{r-R}{\bm n}}}\bm{\nabla}^2 \bm{v}.
	\end{split}
\end{align}

In the fluid region ($r<R$), $\Theta\K{R-r}=1$, $\delta\K{r-R}=0$. Hence, eq.(\ref{TTSH_eq_phi_3}) reduces to
the TTSH equation with no boundaries, eq.(\ref{TTSH_eq_fluid1}).
In the solid region ($r>R$), $\Theta\K{R-r}=0$, $\delta\K{r-R}=0$. 
Thus, eq.(\ref{TTSH_eq_phi_3}) reduces to
\begin{align}
	\label{TTSH_eq_solid_2}
	\partial_{t}\bm{v}=0.
\end{align}
Therefore, by preparing the initial condition where $\bm{v}=0$ in the solid region, $ \bm{v}=0 $ is satisfied at the subsequent time steps.
The equation at the fluid-solid boundary is obtained by integrating eq.(\ref{TTSH_eq_phi_3}) with respect to $r$ from $R-\epsilon$ to $R+\epsilon$, where
$\epsilon>0$.
\begin{align}
	\begin{split}
		\label{TTSH_eq_phi_interface}
		\int_{R-\epsilon}^{R+\epsilon}& dr \partial_t {\bm v}\\
		= &\int_{R-\epsilon}^{R+\epsilon} dr\Theta\K{R-r}\CK{
			-\bm{\nabla}q -\K{\alpha + \beta \ABS{\bm v}^2}\bm{v} \right. \\
			& \left. \quad\quad\quad\quad\quad\quad\quad\quad+\Gamma_0 \bm{\nabla}^2 \bm{v} - \Gamma_2 \K{\bm{\nabla}^2}^2 \bm{v}\right. \\
			& \left. \quad\quad\quad\quad\quad\quad\quad\quad- \lambda_0 \bm{v}\cdot\bm{\nabla}\bm{v}}\\
		&+\int_{R-\epsilon}^{R+\epsilon} dr\delta\K{r-R}\CK{ {\bm n}q - \Gamma_0 {\bm n}\cdot\bm{\nabla} \bm{v}\right. \\ 
			& \left. \quad\quad\quad\quad\quad\quad\quad\quad\quad+ 2 \Gamma_2 \bm{n} \cdot \bm{\nabla} \bm{\nabla}^2 {\bm v}\right. \\
			& \left. \quad\quad\quad\quad\quad\quad\quad\quad\quad-\xi \K{\bm{n}\cdot\Tensor{\epsilon}}\K{\bm{n}\times\bm{v}}_{z}}\\
		&+\int_{R-\epsilon}^{R+\epsilon} dr\Gamma_2 \CK{\bm{\nabla}\cdot\K{\delta\K{r-R}{\bm n}}}\bm{\nabla}^2 \bm{v}
	\end{split}
\end{align}
The l.h.s. and the first term of the r.h.s. of eq.(\ref{TTSH_eq_phi_interface}), whose integrand has a finite value, go to zero
as $\epsilon\rightarrow 0$.
It is quite easy to execute the integral of the second term of the r.h.s..
The third integral of the r.h.s. can be evaluated as follows:
\begin{align}
	\begin{split}
		\label{partial_integral}
		\int_{R-\epsilon}^{R+\epsilon}&dr~ \Gamma_2 \CK{\bm{\nabla}\cdot\K{\delta\K{r-R}{\bm n}}}\bm{\nabla}^2 \bm{v}\\
		=& \int_{R-\epsilon}^{R+\epsilon}dr~ \Gamma_2 \CK{
			\F{1}{r} \PD{}{r}\K{r\delta\K{r-R}}
			}\bm{\nabla}^2 \bm{v}\\
			=& \int_{R-\epsilon}^{R+\epsilon}dr~ \Gamma_2 \PD{}{r} \CK{r \delta\K{r-R} \F{1}{r} \bm{\nabla}^2 \bm{v}}\\
			&- \int_{R-\epsilon}^{R+\epsilon}dr~ \Gamma_2 r\delta\K{r-R} \PD{}{r}\K{\F{1}{r}\bm{\nabla}^2 \bm{v}}\\
			=& - \int_{R-\epsilon}^{R+\epsilon}dr~ \Gamma_2 \delta\K{r-R}\CK{
				- \F{1}{r} \bm{\nabla}^2 \bm{v} + \PD{}{r} \bm{\nabla}^2 \bm{v}}\\
			=& \Gamma_2 \F{1}{R} \bm{\nabla}^2 {\bm v} |_{r=R} - \Gamma_2 {\bm n}\cdot\bm{\nabla} \bm{\nabla}^2 \bm{v} |_{r=R}.
	\end{split}
\end{align}
Therefore, eq.(\ref{TTSH_eq_phi_interface}) reduces to
\begin{align}
	\begin{split}
		\label{TTSH_eq_phi_interface_2}
		0 =& {\bm n}q
		- \Gamma_0 {\bm n}\cdot\bm{\nabla} \bm{v} + \Gamma_2 \bm{n} \cdot \bm{\nabla} \bm{\nabla}^2 {\bm v}\\
		&-\xi \K{\bm{n}\cdot\Tensor{\epsilon}}\K{\bm{n}\times\bm{v}}_{z}
		+ \Gamma_2 \F{1}{R} \bm{\nabla}^2 {\bm v}
		\quad\text{for}\quad r=R.
	\end{split}
\end{align}
To obtain the condition on the tangential conponent, taking the cross product of eq.(\ref{TTSH_eq_phi_interface_2}) with $\bm{n}$, we obtain
\begin{align}
	\begin{split}
		\label{TTSH_eq_phi_interface_3}
		-\xi\K{{\bm n} \times {\bm v}}_z =\BK{{\bm n}\times\K{\Tensor{\sigma}\cdot\bm{n}-\Gamma_{2}\F{1}{R}\bm{\nabla}^2 {\bm v}}}_z \\
		\text{for}\quad r=R.
	\end{split}
\end{align}
This is the Navier slip boundary condition with a correction term
$-\K{\Gamma_2/R}\bm{n}\times\bm{\nabla}^2\bm{v}$.

By taking the dot product of eq.(\ref{TTSH_eq_phi_interface_2}) with ${\bm n}$, we can obtain one more boundary condition:
\begin{align}
	\begin{split}
		\label{TTSH_eq_phi_interface_4}
		0=&q-\Gamma_0{\bm n}\cdot{\bm\nabla}\K{{\bm n}\cdot{\bm v}}\\
		&+\Gamma_2{\bm n}\cdot{\bm\nabla}\K{{\bm n}\cdot{\bm\nabla}^2{\bm v}}+\Gamma_2\F{1}{R}{\bm n}\cdot{\bm\nabla}^2{\bm v}\quad\text{for}\quad r=R.
	\end{split}
\end{align}
In the previous studies using the TTSH equation (ref.\cite{Reinken_2020,Reinken_2022,shiratani2023route}), the zero-vorticity boundary condition ($\omega=0$) is imposed
in addition to the zero-velocity condition (${\bm v}=0$).
Eq.(\ref{TTSH_eq_phi_interface_4}) can be regarded as a boundary condition for the derivatives of $\bm{v}$.
Note that zero-vorticity boundary condition, together with the zero velocity, also corresponds to imposing a boundary condition for the derivative of $\bm{v}$.
Adding an appropriate surface term regarding the vorticity to the functional $\mathscr{F}$ yields a boundary condition imposed directly on the vorticity.
The introduction of such boundary conditions could be the subject of future study.
\section{Stream-function representation of the TTSH equation}
\label{psi_representation}
In our calculations, the basic equation is rewritten
in terms of the stream function $\psi$, defined by $v_i = \epsilon_{ij}\partial_j\psi$.
Substituting $v_i = \epsilon_{ij}\partial_j\psi$ into eq.(\ref{TTSH_eq_phi_2}) and
operating $\epsilon_{ik}\partial_k$ on both sides of the equation, we obtain the
stream-function representation of the TTSH equation:
\begin{align}
	\label{eq:ttsh_psi_phi}
	\begin{split}
		\partial_t\bm{\nabla}^2\psi
		=&-\alpha\CK{\K{\bm{\nabla}\phi}\cdot\bm{\nabla}\psi-\phi\omega}\\
		&-\beta\left\{\K{\bm{\nabla}\phi}\cdot\K{\bm{\nabla}\psi}{\bm v}^2\right.\\
		&\hspace{23pt}\left.+2\phi\K{\bm{\nabla}\psi}\cdot\K{\bm{\nabla}\bm{v}}\cdot{\bm v}
		-\phi{\bm v}^2\omega\right\}\\
		&+\Gamma_0\left\{\K{\bm{\nabla}\bm{\nabla}\phi}:\K{\bm{\nabla}\bm{\nabla}\psi}\right.\\
		&\hspace{25pt}\left.-2\K{\bm{\nabla}\phi}\cdot\bm{\nabla}\omega-\phi\bm{\nabla}^2\omega\right\}\\
		&+\Gamma_2\left\{
			\K{\bm{\nabla}\bm{\nabla}^2\phi}\cdot\bm{\nabla}\omega
			+\K{\bm{\nabla}^2\phi}\K{\bm{\nabla}^2\omega}\right.\\
			&\quad\quad\quad\left.+2\K{\bm{\nabla}\bm{\nabla}\phi}:\K{\bm{\nabla}\bm{\nabla}\omega}
			\right.\\
			&\quad\quad\quad\left. +3\K{\bm{\nabla}\phi}\cdot\K{\bm{\nabla}\bm{\nabla}^2\omega}
			+\phi\K{\bm{\nabla}^2}^2\omega
			\right\}\\
			&+\xi\left\{
				\ABS{\bm{\nabla}\phi}^{-3}\K{\bm{\nabla}\phi}\cdot\K{\bm{\nabla}\bm{\nabla}\phi}
				\cdot\K{\bm{\nabla}\phi}\K{\bm{\nabla}\phi}\cdot\bm{\nabla}\psi
				\right.\\
				&\hspace{25pt}\left.
				-\ABS{\bm{\nabla}\phi}^{-1}\K{\bm{\nabla}\phi}\cdot\K{\bm{\nabla}\bm{\nabla}\phi}
				\cdot\bm{\nabla}\psi\right.\\
				&\hspace{25pt}\left.
				-\ABS{\bm{\nabla}\phi}^{-1}\K{\bm{\nabla}\phi}\cdot\K{\bm{\nabla}\bm{\nabla}\psi}
				\cdot\bm{\nabla}\phi
				\right.\\
				&\hspace{25pt}\left.
				-\ABS{\bm{\nabla}\phi}^{-1}\K{\bm{\nabla}^2\phi}\K{\bm{\nabla}\phi}
				\cdot\bm{\nabla}\psi
				\right\}\\
				&-\lambda_0\left\{\K{\bm{\nabla}\phi}\cdot\K{\bm{\nabla}\bm{\nabla}\psi}\cdot{\bm v}\right.\\
				&\hspace{25pt}\left.+\phi\K{\bm{\nabla}{\bm v}}:\K{\bm{\nabla}\bm{\nabla}\psi}-\phi{\bm v}\cdot\bm{\nabla}\omega\right\},
	\end{split}
\end{align}
where we have introduced the following notation:
\begin{equation}
	\Tensor{T}:\Tensor{U}\equiv T_{ij}U_{ij}.
\end{equation}
As already mentioned in Sec.\ref{practical_techniques}, we used a pseudo-spectral method to calculate eq.(\ref{eq:ttsh_psi_phi}).
In this method, the stream function $ \psi $ is expanded in Fourier series:
\begin{equation}
	\label{psi_fourier_expansion}
	\psi = \sum_{\bm k}\hat{\psi}_{\bm k}e^{i{\bm k}\cdot{\bm r}}.
\end{equation}
Substituting eq.(\ref{psi_fourier_expansion}) into eq.(\ref{eq:ttsh_psi_phi}), we obtain
\begin{equation}
	\label{psihat_ode}
	\OD{\hat{\psi}_{\bm k}}{t}=-\F{1}{\bm{k}^2}\int_{0}^{2\pi}\int_{0}^{2\pi}\F{dxdy}{\K{2\pi}^2}\K{\text{r.h.s. of eq.(\ref{eq:ttsh_psi_phi})}}e^{-i{\bm k}\cdot{\bm r}}.
\end{equation}
\section{non-slip limit}
\label{non_slip_limit}
\begin{figure}[btp]
	\begin{center}
		\includegraphics[width=0.50\textwidth]{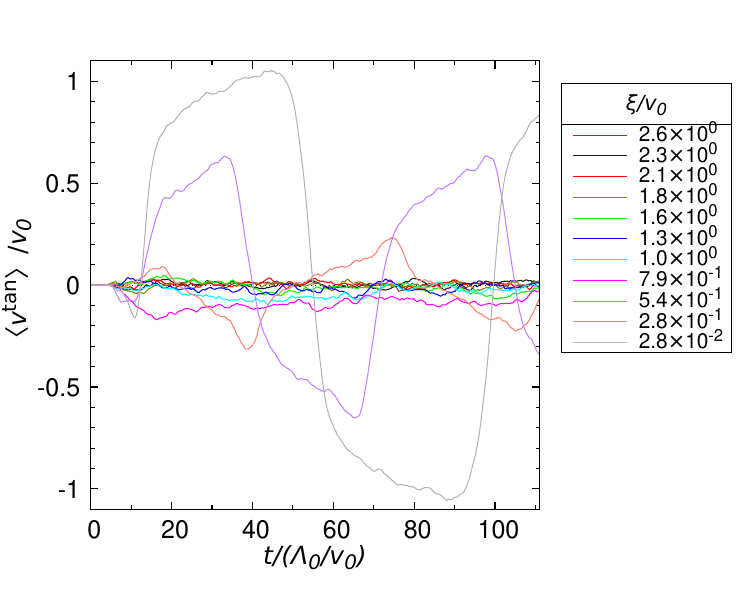}
		\caption{
			Time evolusion of $\braket{v^{\R{tan}}}$ for large values of $\xi$.
			Parameters other than those in TABLE.\ref{table_fixed_parameters} are as follows:
			$h/\K{\Lambda_0/v_0}=0.000555$, $\lambda_0=4.0$ and $R/\Lambda_0=6.3$.
			}
			\label{tmev_v_tan_for_large_xi}
	\end{center}
\end{figure}

To confirm that our model gives the non-slip boundary condition when $\xi\rightarrow\infty$,
we performed simulations in G1 geometry with large values of $\xi$ and typical values of other parameters.
The result (the time evolution of $\braket{v^{\R{tan}}}$) is shown in FIG.\ref{tmev_v_tan_for_large_xi}.
The amplitude of $\braket{v^{\R{tan}}}$ decreases as the drag coefficient $\xi$ increases.
For sufficiently large values of $\xi/v_0$ ($\gtrsim 1.0$), the slip velocity becomes almost zero i.e. the boundary condition reduces to (almost) non-slip.
\section{Threshold value of $\lambda_0$ below which edge current oscillation does not occur}
\label{threshold_lambda_0}
\begin{figure}[btp]
	\begin{center}
		\includegraphics[width=0.50\textwidth]{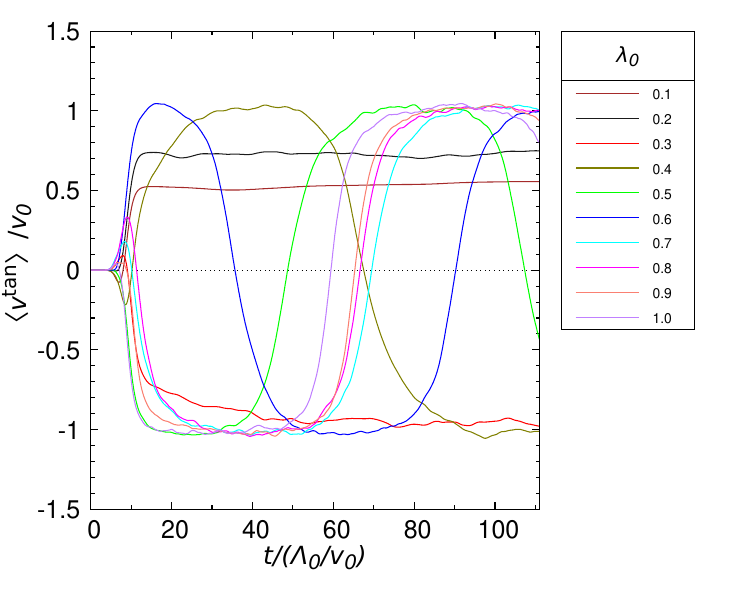}
		\caption{
			Time evolusion of $\braket{v^{\R{tan}}}$ for smaller values of $\lambda_0$.
			Parameters other than those in TABLE.\ref{table_fixed_parameters} are as follows:
			$h/\K{\Lambda_0/v_0}=0.000555$, $R/\Lambda_0=6.3$ and $\xi/v_0=2.8\times 10^{-3}$.
			}
			\label{tmev_v_tan_for_smaller_lambda_0}
	\end{center}
\end{figure}

In Sec.\ref{omega_c_vs_parameters}, we confirmed the absence of edge current oscillation when $\lambda_0=0$ and
its presence when $\lambda_0\geq 2.0$.
Here, one question arises: Is there a finite threshold value of $\lambda_0$, $\lambda_0^{\R{th}}$,
below which the edge current oscillation does not occur?
Let us identify $\lambda_0^{\R{th}}$ for a typical set of parameter values in this Appendix.
To identify $\lambda_0^{\R{th}}$, we performed simulations in G1 geometry, varying the value of $\lambda_0$ from 0.1 to 1.9 at 0.1 intervals.
The result (the time evolution of $\braket{v^{\R{tan}}}$) is shown in FIG.\ref{tmev_v_tan_for_smaller_lambda_0}.
Although we plot the results only for $0.1\leq\lambda_0\leq1.0$ for clarity, we confirmed the oscillation for $1.1\leq\lambda_0\leq1.9$.
From FIG.\ref{tmev_v_tan_for_smaller_lambda_0}, we can identify the threshold value as
\begin{equation}
	\label{lambda_0_threshold}
	0.3 < \lambda_0^{\R{th}} < 0.4.
\end{equation}
\section{The behavior in a narrow channel}
\label{behavior_in_narrow_channel}
In ref.\cite{li2017mechanism,duclos2014perfect}, active nematics are confined in G3 geometry.
In these studies, when the distance between parallel walls is larger than the orientation correlation length,
the particles near the edge align parallel with the wall while particles in the bulk have different directions.
On the other hand, when the distance between walls is equal to or smaller than the orientation correlation length,
the particles in the whole region have the direction parallel with the walls, which is called perfect order.

\begin{figure}[btp]
	\centering
	\includegraphics[width=0.45\textwidth]{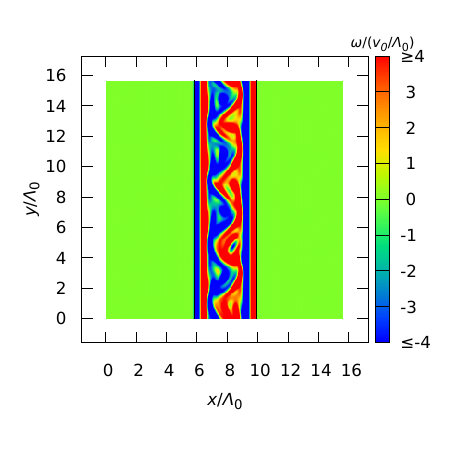}
	\caption{
		A typical simulation snapshot of the vorticity field $ \omega/\K{v_0/\Lambda_0}=\K{\bm{\nabla}\times\bm{v}}_z/\K{v_0/\Lambda_0} $
		at the time $ t/\K{\Lambda_0/v_0} = 274.6 $ for G3 geometry.
		Parameters other than those in TABLE.\ref{table_fixed_parameters} are as follows:
		time increment $ h/\K{\Lambda_0/v_0} = 0.000370$,
		$ \lambda_0 = 4.0 $, $ R/\Lambda_0=1.6 $ and $ \xi/v_0=2.8\times10^{-2} $.
		Note that the spacing of two walls is $2R/\Lambda_0$.
		The black line indicates the outer edge of the smoothed boundary.
		}
		\label{zero_curvature_vorticity_narrow}
\end{figure}

\begin{figure}[btp]
	\centering
	\includegraphics[width=0.45\textwidth]{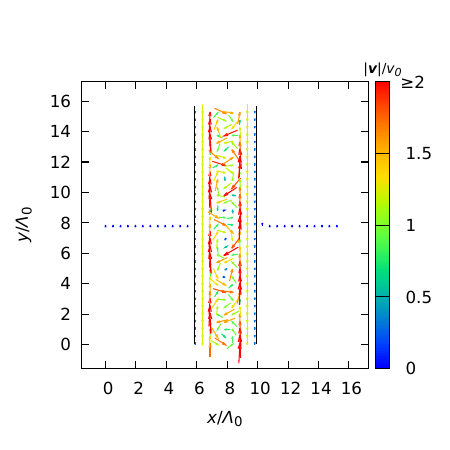}
	\caption{
		A typical simulation snapshot of the velocity field $ \bm{v}/v_0 $ at the time $ t/\K{\Lambda_0/v_0} = 274.6 $ for G3 geometry.
		Parameters other than those in TABLE.\ref{table_fixed_parameters} are the same as in FIG.\ref{zero_curvature_vorticity_narrow}.
		The black line indicates the outer edge of the smoothed boundary.
		Velocity arrows are drawn at intervals of 8 lattice points in each direction.
		}
		\label{zero_curvature_velocity_narrow}
\end{figure}

In our simulations, when the distance between the walls is large, the bulk region exhibits a turbulent behavior
(see FIG.\ref{zero_curvature_vorticity} and \ref{zero_curvature_velocity}).
This behavior is similar to the one in ref.\cite{li2017mechanism,duclos2014perfect}.
When the distance between the walls is small,
vortices line up at regular intervals ($\sim\Lambda_0$)
in the $y$-direction (see FIG.\ref{zero_curvature_vorticity_narrow} and \ref{zero_curvature_velocity_narrow})
and the perfect order observed in the ref.\cite{li2017mechanism,duclos2014perfect} does not emerge.

This difference can be explained as follows.
In the simulations in ref.\cite{li2017mechanism}, the emergence of perfect order is 
attributed to the nematic interaction between the constituent rod-like 
particles, which allows the orientational order near the walls to 
propagate into the channel center. On the other hand, in the TTSH 
simulation, the Swift-Hohenberg term that dictates the typical length 
scale of the spatial pattern destroys the uniform profile of $\bm{v}$, which 
results in the alternate distribution of clockwise and counter-clockwise 
vortices in the $y$-direction. Thus the difference in the behavior of 
order mentioned above is associated with the difference in how the order 
is generated.
\end{appendices}

\end{document}